\documentclass[12pt]{article}
\usepackage{graphics}
\usepackage{amsfonts}
\usepackage{amsmath}
\usepackage{amssymb}

\let\a=\alpha \let\b=\beta  \let\g=\gamma  \let\d=\delta \let\e=\varepsilon
  \let\h=\eta   \let\th=\theta \let\k=\kappa 
\let\m=\mu    \let\n=\nu             \let\r=\rho
\let\s=\sigma \let\t=\tau    \let\ph=\varphi
\let\ps=\psi   
 \let\D=\Delta  \let\L=\Lambda 
    \let\Si=\Sigma     
\let\O=\Omega  \let\del=\nabla

\def\\{\hfill\break} \let\==\equiv

\let\0=\noindent

\let\dpr=\partial

\def\qed{\hfill\raise1pt\hbox{\vrule height5pt width5pt depth0pt}}
      
\let\ra=\rightarrow 

\def\be{\begin{equation}}
\def\ee{\end{equation}}
\def\bea{\begin{eqnarray}}\def\eea{\end{eqnarray}}

\renewcommand{\theequation}{\arabic{section}.\arabic{equation}}
\begin{document}
\markright{Hasen\"ohrl...}

\title{Hasen\"ohrl and the Equivalence of Mass and Energy}

\author {Stephen Boughn{\small\it\thanks{sboughn@haverford.edu}}
~and {Tony Rothman\small\it\thanks{trothman@princeton.edu }}
\\[2mm]
~ \it Haverford College, Haverford PA 19041 \\
~ \it Princeton University, Princeton NJ 08544}

\date{{\small   \LaTeX-ed \today}}

\maketitle

\begin{abstract}
In 1904 Austrian physicist Fritz Hasen\"ohrl (1874-1915) examined blackbody radiation in a reflecting cavity.  By calculating the work necessary to keep the cavity moving at a constant velocity against the radiation pressure he concluded that to a moving observer the energy of the radiation would appear to increase by an amount $E= (3/8)mc^2$, which in early 1905 he corrected to $E=(3/4) mc^2$.  Because relativistic corrections come in at order $v^2/c^2$ and Hasen\"ohrl's {\it gedankenexperiment}  evidently required calculations only to order $v/c$, it is initially  puzzling why he did not achieve the answer universally accepted today.  Moreover, that $m$ should be equal to $(4/3)E/c^2$  has led commentators to believe that this problem is identical to the famous ``4/3 problem" of the self-energy of the electron and they have invariably attributed Hasen\"ohrl's mistake to neglect of the cavity stresses.  We examine Hasen\"ohrl's papers from a modern, relativistic point of view in an attempt to understand where exactly he went wrong. The problem turns out to be a rich and challenging one with strong resonances to matters that remain controversial.   We  give an acceptable relativistic solution to the conundrum and show that virtually everything ever written about Hasen\"ohrl's thought experiment, including a 1923 paper by Enrico Fermi, is misleading if not incorrect.

 \vspace*{5mm} \noindent PACS: 03.30.+p, 01.65.+g, 03.50.De

\end{abstract}
\section{Historical introduction}
\setcounter{equation}{0}\label{sec1}
\baselineskip 8mm

If the name Fritz Hasen\"ohrl has not utterly vanished from the annals of physics it is only because when delving into the early history of relativity, amidst the more indelible names of Einstein, Poincar\'e, Lorentz and Abraham, one occasionally stumbles across mention of the Austrian physicist who in 1904 attempted the first serious calculation to establish the connection between the energy of radiation and inertial mass.  Unfortunately, Hasen\"ohrl  arrived  at $E=(3/8) mc^2$, which he later amended to $E = (3/4)mc^2$.  Because he  was working within the confines of an ether theory, Hasen\"ohrl's incorrect results were soon displaced by Einstein's correct one.  Nevertheless, since relativity corrects ether theories at second order  in $v/c$ and derivations of $E=mc^2$ have been proposed that  require calculations only to first order ---evidently as in Hasen\"orhl's own thought experiment---it is far from obvious why he could not and did not obtain the correct answer.

The idea that mass and energy should be related did not originate with Hasen\"ohrl himself and, to be sure, only in popular mythology did the connection between the two quantities suddenly emerge in the miraculous year of 1905.  Once Maxwell's theory had been established in the 1860s, physicists soon began the attempt to place all natural phenomena on an electromagnetic footing.  As early as 1881 J.J. Thomson\cite{Thom81} argued that the backreaction of the magnetic field of a charged sphere would impede its motion and result in  an apparent mass increase of $(4/15)\mu e^2/a$, where $e$ was the charge on the sphere, $a$ its radius and $\mu$ the magnetic permeability.  His results were criticized and elaborated by Fitzgerald and Heaviside, the latter of whom showed\cite{Heavi89} that the mass increase of a moving sphere with uniform surface charge distribution was $m = (4/3)E_o/c^2$, for stationary field energy $E_o$. (For more on these early researches, see Max Jammer's \emph {Concepts of Mass}\cite{Jammer51}). A similar investigation was carried out by Wien, whose result agreed with Heaviside's\cite{Jammer51}. In 1900 a consideration of Poynting's theorem led Henri Poincar\'e\cite{Poin00} to state that electromagnetic momentum acted as a ``{\it fluide fictif}" with a mass such that field energy $E=mc^2$.  Poincar\'e, however, did no serious calculation.  The most outspoken proponent of the electromagnetic origin of mass was Max Abraham, who in 1903 concluded\cite{Ab03, Jammer51}, in agreement with Heaviside, that due to interaction with its field an electron in motion would have an apparent mass given by $m=(4/3)E_o/c^2$.

All these investigations, with the exception of Poincar\'e's, were confined to what we would today call the mass-energy relationship between the classical electron and its field, whereas in asking for the mass equivalent of blackbody radiation, Hasen\"ohrl broadened the scope of the investigations.  As to why he achieved an incorrect result, available references do not clarify matters.  In his  \emph {Concepts of Mass in Contemporary Physics and Philosophy}\cite{Jammer00}, Jammer, adding a few words to his earlier \emph{Concepts of Mass}, says only: ``What was probably the most publicized prerelativistic declaration of such a relationship [between inertia and energy] was made in 1904 by Fritz Hasen\"ohrl. Using Abraham's theory, Hasen\"ohrl showed that a cavity with perfectly reflecting walls behaves, if set in motion, as if it has a mass $m$ given by $m = 8V\e_0/3c^2$, where $V$ is the volume of the cavity, $\e_0$ is the energy density at rest, and $c$ is the velocity of light."  Nor does Edmund Whittiker provide more insight. ``In 1904 F. Hasen\"ohrl," Whittiker writes in his monumental \emph{A History of the Theories of Aether and Electricity}\cite{Whit87}, ``considered a hollow box with perfectly reflecting walls filled with radiation, and found that when it is in motion there is an apparent addition to its mass, of amount $(8/3c^2)$ times the energy possessed by the radiation when the box is at rest: in the following year he corrected this to $(4/3c^2)$ times the energy possessed by the radiation when the box is at rest..."  Abraham Pais, in his biography of Einstein\cite{Pais82}, accords Hasen\"ohrl only a short footnote that is even vaguer than the previous remarks: before 1905 Hasen\"ohrl showed that kinetic energy of a cavity filled with radiation increases ``in such a way that the mass of the system appears to increase."

The impression given by such statements is that none of the authors has troubled  to understand exactly what Hasen\"ohrl did.  Current online references only reinforce this opinion.  Herrmann\cite{Hermann04} contends, ``Indeed, in 1914, Cummingham [\emph{sic}](1, p. 189) shows that Hasen\"ohrl made a slight error in that the shell is not included in his calculations. If the shell had been included, then the factor would be 1 or $m = E/c^2$."  The Wikipedia entry on Hasen\"ohrl repeats the assertion, adding that he could not have correctly modeled the shell ``since he did not have relativistic mechanics\cite{Wiki1}."  Hasen\"ohrl's case was probably not served by Philipp L\'enard, a virulent antisemite who later became Hitler's chief of Aryan physics\cite{DSBL} and who in 1921 published an article\cite{Lenard21}  attempting to discredit Einstein by giving priority for mass-energy equivalence to Hasen\"ohrl and the gravitational deflection of light to   Bavarian astronomer
Johann Georg von Soldner (1776-1833), even declaring that the inertia of energy be called the ``Hasen\"ohrlsche Masse."  Polemics aside, there was an element of truth in L\'enard's assertions.  In 1801 (\emph {sic}) Soldner, a gifted mathematician, made a prescient calculation of light deflection by the sun, obtaining the correct Newtonian result\cite{Soldner}. And despite Hasen\"ohrl's incorrect result, he did clearly state that blackbody radiation has an apparent mass associated with its velocity.

In attempting to clear up all these matters, we take the unusual approach of examining Hasen\"ohrl's original papers.  To do so is not straightforward.  No translation of them exists and machine translations provide little more than gibberish.  The papers themselves\cite{H1,H2,H3} (henceforth H1, H2 and H3), are by today's standards presented in a cumbersome manner and are, obviously, not free of error.  The greatest hinderance, however, is that they are written from an obsolete world view, which can only confuse the reader steeped in relativistic physics.  Nevertheless, in certain ways Hasen\"ohrl's thought experiment was more audacious than Einstein's and because it is at least superficially related to the famous ``4/3 problem" of the self-energy of the electron (see \S\S 3,4,6), it becomes central to twentieth-century physics.  Enrico Fermi, in fact, assumed that the two 4/3's were identical and devoted one of his earliest papers to resolving the paradox\cite{Fermi23b}.  The overall plan for our exercise in ``forensic physics" is to first introduce Hasen\"ohrl's thought experiment (\S\ref{sec2}), then achieve a correct relativistic result (\S\ref{sec3} and \S\ref{sec5}), which will allow us to understand why his proof failed.  In the process we determine that cavity stresses are irrelevant (\S\ref{sec4}) and that, if Fermi solved anything, it was not entirely Hasen\"ohrl's problem  (\S\ref{sec6}).  We end by comparing Hasen\"ohrl's proof to Einstein's (\S\ref{sec7}).  An appendix gives a few details of Hasen\"ohrl's papers.\\

Before moving on to physics proper, it is worth saying a few words about Hasen\"ohrl himself, since his name is unfamiliar to most physicists.  Fritz Hasen\"ohrl was born on 30 November, 1874 in Vienna\cite{DSBH}.  He later studied  mathematics and physics at Vienna University under Franz Exner and Ludwig Boltzmann. When Hasen\"ohrl received his Ph.D. in 1897, Boltzmann sent ``his best student" to  Leiden, where he spent a year as a research assistant to Kamerlingh Onnes, after which he returned to Vienna as a {\it privatdozent}. Hasen\"ohrl's first systematic research, begun for his Vienna dissertation under Exner and continued in Leiden, was an experimental investigation of the temperature dependence of the dielectric constants of liquids and solids.  In 1904--1905 he published the three papers for which he is best known, ``On the theory of radiation in moving bodies," which concerned  the mass equivalent of blackbody radiation in a moving cavity.  The second and third of these papers (H2, H3) appeared in the \emph {Annalen der Physik} and for his work Hasen\"ohrl won the Haitinger Prize of the Austrian Academy of Sciences. In 1907 he succeeded Boltzmann as professor of physics at Vienna after the latter's suicide.  Hasen\"ohrl counted Hans Thirring and Erwin Schr\"odinger among his students.

Hasen\"ohrl was unquestionably held in high regard.  Of his lectures Thirring recollected, ``Hasen\"ohrl's art of giving lectures was quite excellent. He was the best lecturer I ever heard, much better, for instance, than Wirtinger's lectures. Hasen\"ohrl was the cause for me that I turned to theoretical physics. I first started having more interest in experimental physics but when I heard Hasenohrl's lectures, I was so fascinated by the way he did it that I became a theoretical physicist too\cite{Thirr63}."

 Schr\"odinger, who was a year ahead of  Thirring, himself says, ``No other person has had a stronger influence on me than Fritz Hasen\"ohrl, except perhaps my father.  There was a certain air of chivalry about Hasen\"ohrl, and his friendliness overcame any barriers of formality or seniority between him and his students.  He often had groups of them to his house, where his beautiful wife Ella presided and his small son and daughter added to the happy atmosphere.  He was a strong mountaineer and expert in skiing and other winter sports.  He organized expeditions with the students and took an interest in student affairs and, as Hans Thirring reported, wherever he went he acted as an energizer and brought good fellowship\cite{Moore89}."

Hasen\"ohrl's lectures covered the foundations of analytical mechanics, electromagnetic theory, optics and statistical mechanics.  At the outbreak of World War I, he joined the Austro-Hungarian  army and was killed in the Isonzo campaign.

\section{Hasen\"ohrl's thought experiment}
\setcounter{equation}{0}\label{sec2}

\begin{figure}[htb]
\vbox{\hfil\scalebox{.6}
{\includegraphics{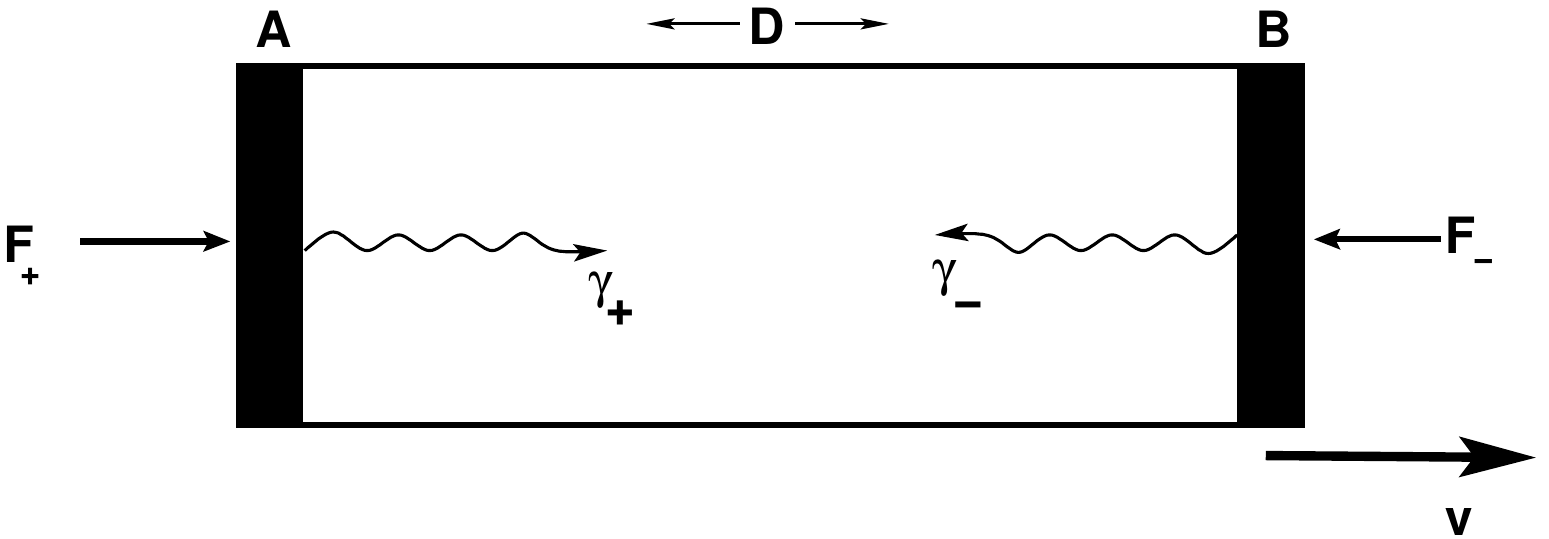}}\hfil}
{\caption{\footnotesize{A cavity consisting of two blackbody radiators, $A$ and $B$ in a completely reflecting enclosure of length $D$.  At a time $t=0$ the radiating caps suddenly begin to emit photons in the direction of motion (+) and opposite the direction of motion (-).  From the frame of a moving observer, the (+/-) photons will be blue/red shifted and hence exert different reaction forces on $A$ and $B$. \label{cav1}}}}
\end{figure}

Given the primacy of blackbody radiation in 1900, Hasen\"ohrl's thought experiment was both natural and reasonable.  Unfortunately, it turned out not to be simple.  He considered  two blackbody radiators $A$ and $B$ placed inside a cavity with perfectly reflecting walls.  He assumes that the region exterior to the cavity is at absolute zero temperature and that the cavity walls surround the radiators' outer surfaces, such that any emitted radiation is directed inward (see figure \ref{cav1}).  At a time $t=0$, $A$ and $B$ are turned on, filling the cavity with radiation, and at the same instant external forces $F_\pm$ are applied to keep the cavity motionless.  Clearly, in the rest frame of the cavity, nothing whatsoever happens regarding the cavity's motion.  The situation is different, however, for an observer moving with respect to the cavity; we refer to this observer's frame as the lab.

We should point out immediately that Hasen\"ohrl does not use the terminology ``rest frame of the cavity" or ``lab frame."  Although he mentions the ether only three times throughout his papers, it is clear that for him all motion is taking place relative to some absolute reference frame.  In particular, the cavity is really moving through the lab at an absolute velocity $v$.  For the body of this paper we abandon that viewpoint and speak of the lab or cavity frames with their modern meanings.  \emph{We designate important quantities in the rest frame of the cavity, such as energy, by a subscript $o$. Other rest-frame quantities are unprimed.  Quantities referring to the lab frame are designated with a prime ($\ '$).}

To give the basic idea of how Hasen\"ohrl's experiment is to meant to operate, consider first a toy version, in which we assume that relativity is correct, in particular that the speed of light $c$ is constant.  Suppose that the lab observer sees the cavity moving at constant velocity $v$ in the positive $x$-direction, as shown in figure 1.  This observer sees the left radiator $A$ emitting photons, which we  take to be plane waves  emitted  to the right.  They  therefore exert a reaction force to the left.  But according to the lab observer the external force $F'_{ +} $  acting against the radiation recoil is such as to keep $A$ moving at constant velocity.  The work done against the radiation pressure in a time $dt'$ will be simply
\be
dW'_+ = F'_+ v dt' = A_o p'_{\g+}v dt',
\ee
where $p'_{\g +}$ is the radiation pressure exerted on the left radiator as observed in the lab and $A_o$ is the cross-sectional area of the cavity, which is the same in both frames.  Similarly, the lab observer will ascribe a force $F'_{-}$ acting to the left and resulting in work
\be
dW'_- = F'_- v dt' = -A_o p'_{\g-}v dt'.
\ee
The net work done in $dt'$ will then be
\be
\D W' = (p'_{\g+} - p'_{\g-}) A_o v dt'. \label{DW1}
\ee

We know that in the lab the right-moving light will be Doppler blue-shifted and the left-moving light will be Doppler red-shifted.  Relativistically, the photon pressure $p'_{\g} = i'/c$ for intensity $i' =$ (\emph {energy per unit time per unit area}), will be given by
\be
p'_{\g \pm}= p_o\frac {1 -\b^2} {(1\mp \b)^2},
\ee
with $\b \equiv v/c$.  For $v << c$ we have $p'_{\g \pm} \approx p_o(1\pm 2\b)$.  (Intensity, as opposed to frequency, is Doppler shifted by two powers of $1\pm \b$.  Heuristically, since $E=h \n$, the energy of each photon is Doppler shifted by one power of $1\pm \b$, but the number of photons passing an observer per unit time is increased by the same factor, so the  intensity, which is $\sim n\n$, gets two powers of $1\pm \b$.  The same is not true for isotropic radiation; see \S\ref{subsec3.2} )

What $dt'$ are we to use in Eq. (\ref{DW1})?  After one light-crossing time, blueshifted photons from $A$ will be absorbed by $B$ and, conversely, redshifted photons from $B$ will be absorbed by $A$.  Thereafter, $F'_+ = F'_-$ are equal and the total work done is thus  zero.  Hasen\"ohrl recognizes this (H1, H2) and, consequently, carries out his computation for only one light-crossing time, which to the required order in $\b$  in the toy model is just $D/c$.   Eq. (\ref{DW1}) then becomes (to lowest order in $\b$)
\be
\D W' = p_oD A_o [(1+2\b)  - (1-2\b)]\frac{v}{c} = 4p_o D A_o \b^2. \label{DW2}
\ee
However, for plane waves $p_o = \r_o = $ (\emph{energy per unit volume}), and so $p_o D A_o = \r_oDA_o$ is merely the total rest energy of the radiation, $E_o$.  Therefore  $\D W' = 4E_o\b^2$.   Hasen\"ohrl now notices that this expression resembles a kinetic energy.  He  assumes that the radiation has an equivalent mass $m$ and uses the work-energy theorem to equate the net work done to the increase in  kinetic energy of this mass.  Then one has immediately
\be
4E_o \frac{v^2}{c^2} = \frac1{2} m v^2,
\ee
or
\be
E_o = \frac1{8} mc^2.
\ee

This toy model gives the essence of Hasen\"ohrl's calculation in his June 1904 paper, H1.  Notice that we only needed to calculate the pressure to first order in $v/c$ in order to determine the work to second order in $v/c$.  Complications such as the time dilation of $dt$ and the Lorentz contraction of $D$ would appear to be irrelevant, and for this reason we said in the Introduction that it is initially puzzling that Hasen\"ohrl did not achieve the right answer.  Notice also that we said nothing about the mass of the cavity.   This is as in Hasen\"ohrl.  He makes no mention of cavity properties, apart from the fact that the walls are reflecting.  In order to calculate at constant $v$, however, he does assume that at the  moment the radiators are switched on, the cavity velocity instantaneously becomes $v$.  Of course this can only be the case if the mass of the cavity is zero (see also \S\ref{subsec5.2}, below).  The assumption does not appear necessary; the calculation goes through unchanged if the cavity is already moving at $v$ at the instant the radiators are turned on.

The principle differences between the toy model and Hasen\"orhl's more elaborate one are that: 1) he did not assume $c$ is constant; 2) he did not make use of the Doppler shift, although in H1 and H2 he states that a (classical) Doppler-shift calculation gives the same results as his own; 3) instead of plane waves, he takes the more realistic case of blackbody radition, which is emitted isotropically from the end caps, and which requires him to integrate the pressure over angle. One might already guess that averaging over directions will introduce a factor of three in the above answer, leading to a work $\D W' = (4/3) E_o\b^2$ and Hasen\"orhl's  result in H1 of $E_o = (3/8) mc^2$, which allows him to state that an``apparent mass of $(8/3) E/c^2$" is added to the energy of the radiation due to the motion of the cavity.  In his July 1904 paper, H2, he considers a slowly accelerating cavity for which, modulo an algebraic error, he gets $E_o = (3/4) mc^2$. As the result of a communication from Abraham, the error is brought to light and corrected in his 1905 paper H3.  We now derive these results  rigorously  as we attempt to discover what went awry through the  construction of a fully relativistic model.

\section{Relativistic considerations}
\setcounter{equation}{0}\label{sec3}

\subsection{Energy-momentum tensor}
\label{subsec3.1}

We begin by considering the constant-velocity case of H1.  The most straightforward way to obtain the expected answer is to transform the energy-momentum tensor for the radiation.  The energy-momentum tensor for blackbody radiation is the same as for a perfect fluid with equation of state $p=(1/3)\r$ and has the form
\be
T_{\m\n} = \frac1{c^2}(\r + p)u_\m u_\n + \h_{\m\n}p. \label{Tuv}
\ee
Here, all the symbols have their usual meanings:  ${\bf u} \equiv (\g c , \g \bf v)$ is the four velocity, $\bf v$ is the three-velocity, $\g \equiv (1 - \b^2)^{-1/2}$ and the metric tensor $\h_{\m\n} \equiv (-1,+1,+1,+1)$.  Greek indices range from 0...3 and Latin indices take on the values 1...3.  Thus, if $\r_o$ and $p_o$ represent the density and pressure of radiation in the rest frame of the cavity, then in the lab frame
\be
T'_{0 0} = (\r_o + p_o)\g^2 -p_o =  \r_o\g^2 + \frac{\r_o}{3}(\g^2 - 1),\label{T00}
\ee
and
\be
T'_{0 x} = (\r_o + p_o)\g^2\frac{v}{c}  = \frac{4}{3}\r_o\g^2\frac{v}{c}, \label{T0k}
\ee
where  $v = v_x$.

Because $T_{00}$ represents energy density, the total energy in the lab frame will be
\be
E' = \int T'_{00} dV'  \label{E'1}
\ee
where $dV'$ is the volume element in the lab.  Now, the Lorentz contraction is in the $x$-direction only, and hence we take
\be
dV' = \g^{-1}dV_o.  \label{dV'}
\ee
Therefore,
\be
E' = \g^{-1}T'_{00}V_o \label{E'2}
\ee
and from Eq. (\ref{T00}),
\be
E' =  \g E_o(1+ \frac{\b^2}{3}) = E_o(1+\frac{5}{6}\b^2) + {\cal O}(\b^4). \label{E'3}
\ee

 Similarly, from Eq. (\ref{T0k}), the total momentum of the radiation in the lab frame will be
\be
G' = \frac1{c} \int T'_{0x} dV',
\ee
or
\be
G' =\frac{4}{3}E_o\g \frac{v}{c^2} = \frac{4}{3}E_o \frac{v}{c^2} + {\cal O}(\b^3), \label{G'}
\ee
where we have reverted to the old symbol for momentum, $G$, so as not to confuse it with  pressure.

In carrying out this simple derivation we are unexpectedly confronted by  a serious dilemma. If we regard the external force $F$ in the previous section as the time derivative of the momentum, then we  expect from Eq. (\ref{G'})
\be
\D W' = F'_{net} v dt' = \D G' v \approx \frac{4}{3}E_o \frac{v^2}{c^2},
\ee
which, as we argued in \S\ref{sec2}, is the value that should be anticipated from an integration over angle, and which via the work-energy theorem does lead to Hasen\"ohrl's initial result $m = (8/3)E_o/c^2$.  Independently of Hasen\"ohrl, however, $\D W' = (4/3)E_o \b^2$ is \emph {not} equal to $\D E' = (5/6) E_o \b^2$ obtained from Eq. (\ref{E'3}).  If we have put $(4/3)\b^2E_o$ worth of work into a system of pure radiation, it is hardly clear where else it can go other than into increasing the energy of the photons, which is precisely the change $\D E' = (5/6) E_o \b^2$.  Does $(4/3)\b^2E_o$ include something other than kinetic energy?  Has $(1/2)\b^2E_o$ worth of energy vanished into the ether?

This incompatibility is apparently a reflection of an even more profound difficulty.  Whereas elementary texts define the energy-momentum four vector $ {\bf g} = (\g m_o c, \g m_o{\bf  v})$ in order that $g_\n g^\n = -m_o^2c^2 = -E^2_o/c^2$, we see that $E'$ in Eq. (\ref{E'3}) and $G'$ in Eq. (\ref{G'}) \emph{do not} form a four vector.  That is, $-E'^2 + G'^2 \ne E_o^2/c^2$, which means that either $E_o$ is not a Lorentz invariant or that we have inadvertently performed some non-covariant operation.  The difference between the elementary situation and the present one is that we are here dealing with an extended object, which historically have provided numerous paradoxes and controversies in relativity.  Mathematically the debatable step lies in Eq. (\ref{dV'}), which we have employed as the volume element in the spatial integration of Eq. (\ref{E'1}).  To calculate the total energy in the cavity rest frame, we integrated the energy density over the volume at  \emph{constant time}.  But due to the relativity of simultaneity, such a constant-time hypersurface must correspond to events that take place at different times in a Lorentz-transformed  frame.

 It is precisely this quandary that emerges in the famous ``$4/3$ problem" of the energy of the electron. The classical electron, an elastic sphere of radius $a$ with charge $e$ uniformly distributed on the surface, has an external electromagnetic field energy
 \be
(U_o)_{ em} = \int_a^\infty T_{00} dV , \label{Uo}
\ee
where $T_{00}= (1/8\pi)(E^2 + B^2)$  is the energy density of the electromagnetic field.  In the rest frame of the electron $B=0$ and the electrostatic field $E = e/r $.  Hence the result is $(U_o)_{ em} =  e^2/2a$.  Under a Lorentz transformation in the $x$-direction, however, the field components transform as
\bea
E'_x = E_{ox}  & E'_y = \g E_{y}  & E'_{z} = \g E_z \nonumber\\
B'_x = 0     & B'_y = -\b\g E_{z} & B'_z = \b\g E_{y}
\eea
Upon performing the integration  in Eq. (\ref{E'1}) with $dV'$ as in Eq. (\ref{dV'}), we get exactly Eq. (\ref{E'3}), that is, $U'_{em} =\g (U_o)_{ em} (1+\b^2/3)$.  On the other hand, if one integrates the electromagnetic momentum density
\be
G'_{em}= \frac1{4\pi c}\int ({\bf E' \times B'}) dV',
\ee
then one gets $G'_{em} = (4/3)\g G_o v/c$, exactly Eq. (\ref{G'}).  This is the ``4/3 problem," although most papers discussing it evaluate only $G_{em}$, not $U_{em}$, leaving the impression that the coefficient to $U'_{em}$ should be unity, which is not the case.

Given that both the classical electron and Hasen\"ohrl's  experiment result in exactly the same discrepancy, it is not surprising that one might view them as being the same problem, as did Fermi\cite{Fermi23b}.  We will return to a fuller discussion of this issue in \S\S\ref{sec4}--\ref{sec5}. For the moment we assert that, as perplexing as they might seem, the expressions for $E'$ and  $G'$ are correct, if one accepts the integration procedure as valid.  Of course, Hasen\"ohrl in 1904 did not have relativistic tensor methods at his disposal.  He made a dynamical calculation of the work done on the radiation, a more elaborate version of the toy model in \S\ref{sec2}.  To uncover his error we want to make a similar but relativistically correct calculation, which will allow a direct comparison with his results.

\subsection{Relativistic calculation of the work}
\label{subsec3.2}

Our strategy is to take a photon picture and  calculate the radiation pressure on a moving surface by transforming the radiation intensity $i_o$ in the cavity rest frame to a frame moving at an arbitrary velocity relative to the cavity.  Because we are now considering intensity as a function of angle for a given frequency, we must take into account 1) the relativistic Doppler shift; 2) the abberation of the angle of emission from the radiators due to the motion of the detector; 3) the transformation of the volume element due to abberation; 4) time dilation.  We follow the method outlined by Peebles and Wilkinson\cite{PW68}(henceforth PW) for a calculation of the anisotropy of the cosmic microwave background radiation, although same transformation laws can be found in Pauli's 1921 \emph{Relativit\"atstheorie}\cite{Pauli}.

\begin{figure}[htb]
\vbox{\hfil\scalebox{.6}
{\includegraphics{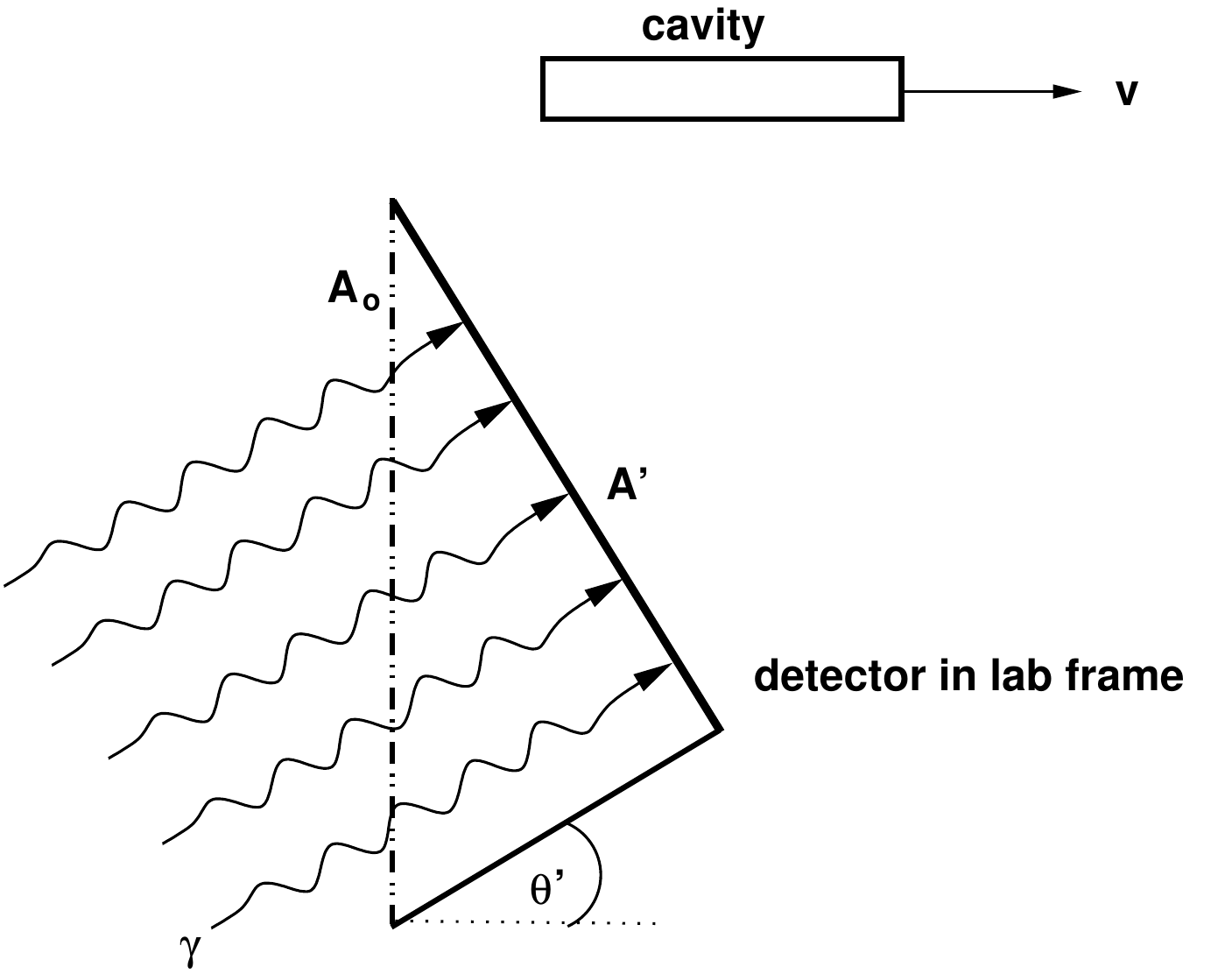}}\hfil}
{\caption{\footnotesize{Photons enter a detector in the lab at an angle $\th'$ to be detected on surface $A'$.  To reach $A'$ the photons must pass through a surface $A_o$, whose normal is aligned with $\bf v$, the cavity's velocity with respect to the lab. Note that $A'=A'_o\cos\th'$.\label{pw}}}}
\end{figure}

Let an observer in the lab frame carry a detector with a collecting area $A'$.  Photons  leaving, say, the cavity's left wall at an angle $\th'$,  enter the instrument and strike $A'$, which responds to photons entering within a solid angle $d\O'$ of the normal to $A'$ (see figure \ref{pw}). In order to be detected the photons must pass through an area $A_o'$ whose normal is chosen to point in the direction the velocity $\bf v$ of the  cavity, which is  to the right in the lab.   Note that $A' = A_o' \cos\th'$ and that $A_o' = A_o$, since there is no Lorentz contraction in directions perpendicular to $\bf v$.  In a time $dt'$,  photons sweeping out a volume $c\cdot A_o \cdot \cos\th'\, dt'$ will strike the detector surface $A'$.  If there are $dN' = n' V'$ photons in this volume, then
\be
dN' = n'(\th')\; d\O' \; A_o'\; c \cdot \cos\th'\; dt',
\ee
and $n'(\th')\equiv$ \emph{number of photons per unit volume per unit solid angle}.  (Unlike PW, we assume monochromatic radiation.)

In the cavity's rest frame, an observer will see the detector moving left and the relative $x$-velocity of the photons and cavity will be $c\cdot \cos\th + v$.  Thus, if this observer reads off from the detector the same number of photons, they will have come from the larger volume $ A_o (c\cdot \cos\th + v)dt$.  Hence
\be
dN = n(\th) \; d\O \; A_o \; ( c\cdot \cos\th + v) \; dt ,
\ee
But since $dN  = dN'$ and $A_o' = A_o$,
\be
n' = n  \frac{d\O}{d\O'}\; \frac{(\cos\th  + \b)}{\cos\th'}\; \frac{dt }{dt'}. \label{n'}
\ee

Now, in the photon picture, the intensity in the lab frame is $i' = n' c h \n'$, while in the cavity frame $i_o = n  c h \n$. (Note that we measure $i'$ and $i_o$ in \emph {energy per unit area per unit time per unit solid angle}).  Therefore from Eq. (\ref{n'})
\be
i' = n  \frac{d\O }{d\O'} \frac{(\cos\th  + \b)}{\cos\th'} \frac{dt }{dt'}\; h\n' \label{i'}
\ee
All that remains is evaluate the various factors on the right hand side of this equation.  The value of $\n'$ is given by the relativistic Doppler shift:
\be
\n' = \left[\frac{1+\b\cos\th }{(1-\b^2)^{1/2}}\right] \n,  \label{dop}
\ee
The time dilation factor is of course
\be
\frac{dt}{dt'} = \g = \frac1{(1-\b^2)^{1/2}}.
\ee
The transformation of the angles is given by relativistic aberration:
\be
\cos\th' = \frac{\cos\th  + \b}{1+\b\cos\th } , \label{costh'}
\ee
and
\be
\frac{d\O'}{d\O} = \frac{d(\cos\th')}{d(\cos\th)}=\frac{(1-\b^2)}{(1+\b\cos\th)^2} .  \label{dO'}
\ee
Inserting all these factors into Eq. (\ref{i'}) yields
\be
i' = i_o \frac{(1+\b\cos\th )^4}{(1-\b^2)^2}.  \label{i'th}
\ee

This expression, however, gives the intensity in the lab frame in terms of the angles in the cavity frame.  It is useful to have an expression for $i'$ entirely in terms of primed quantities.  To write Eq. (\ref{i'th}) in terms of $\th'$ we need the inverse transformations of Eqs. (\ref{dop}--\ref{dO'}). These are  obtained merely by swapping primed and unprimed variables and letting $\b \ra -\b$.  One then easily finds
\be
i' = i_o \frac{(1-\b^2)^2}{(1-\b\cos\th')^4}. \label{i'th'}
\ee
Eq. (\ref{i'th}) or Eq. (\ref{i'th'}) provides the basis for a calculation of the work in our analysis. Note that to calculate  $\D W'$ to first order in $\b$ as we did in $\S\ref{sec2}$, one can ignore the factors of $(1-\b^2)$ in Eq. (\ref{i'th}), which are relativistic corrections, and simply use the  classical limits of the above factors in  Eq. (\ref{i'}).

We further point out that if by blackbody radiation one means isotropic radiation with a Planckian spectrum, then even if the radiation is isotropic in the cavity frame ($i_o = constant$), it cannot be blackbody radiation in the lab frame, since it has become anisotropic.  (The  spectrum does remain Planckian in each direction $\th'$ with a Doppler-shifted temperature; see PW.)  If that is the case, one might question the very basis of Hasen\"ohrl's experiment since he is not employing blackbody radiation at all.  Hasen\"ohrl, however, invokes no properties of the radiation other than its isotropy in the cavity frame, in which case $i_o$ is the specific intensity and the energy radiated into $d\O$ through a surface area element $d\s$ in a time $dt $ is $d{ E_o } = i_o \cos\th \, d\O \,  d\s \, dt$, where the $\cos\th$ is from Lambert's law.  Integrating over angle gives $ dE_o(\s,\t) = \pi i_o d\s  \, dt$.  If $d\s$ is radiating into a small volume at distance $r$ with solid angle element $dA/r^2$, it  is then not difficult to show\cite{Planck12} that the energy density in a cavity filled with isotropic radiation is
\be
\r_o = \frac{4\pi i_o}{c},  \label{ro}
\ee
which Hasen\"ohrl accepts (H1).\\

We are now in the position to compute the work in a way analogous to Hasen\"ohrl's.  One can do this with quantities from either frame; we chose to use all lab-frame variables.  Consider the radiation being emitted at an angle $\th'$ from the left wall of the cavity in figure 1.  Then
\be
\frac{dE'}{dt'} = i' \; d\O' A_o \; (\cos\th' - \b),  \label{dEdt}
\ee
where the last factor is due to the relative velocity of the photons and the wall.  We need the normal component of the momentum.  Since for photons $g_\g = E_\g/c$,
\be
\frac{dg'_x}{dt'} = \frac{i'}{c} \; d\O' A_o \; (\cos\th' - \b)\cos\th'.
\ee
From Newton's third law, the leftward reaction force exerted by the photons on the wall has this magnitude and if the wall is to remain at constant velocity then an equal and opposite external force directed rightward must be applied to the wall.  Following Hasen\"ohrl, the element of work done by that portion of the external force balancing the reaction force of the radiation emitted at an angle $\th'$ from the left wall will be
\be
dW_+ (\th')= \frac{dg'_x}{dt'} \cdot dx = \frac{dg'_x}{dt'} \cdot v \cdot dt'({\th'}),
\ee
where we take $dt'({\th'})$ to be the light-crossing time for a photon at an angle $\th'$.  From elementary geometry it is easy to show that, independent of the number of reflections along the cavity wall,
\be
dt'(\th') = \frac{D'}{c(\cos\th' - \b)}.  \label{dt'}
\ee
For cylindrical symmetry $d\O' =  2\pi\sin\th' d\th' = -2\pi d(\cos\th')$ and hence
\be
dW_+ (\th') = -2\pi i_o \frac{A_o D' v (1-\b^2)\cos\th' d(\cos\th')}{c^2(1-\b\cos\th')^4}, \label{dW+}
\ee
Retaining terms in the integrand to first order in $\b$, the work on radiator $A$ is therefore
\bea
\D W_+ &=& \frac{2\pi i_o A_o D v}{c^2} \int_0^1 (1+4\b\cos\th')\cos\th' d(cos\th')\nonumber\\
      &=&  \frac{2\pi i_o A_o D v}{c^2} \left(\frac1{2} + \frac{4\b}{3}\right).
\eea
It is not difficult to convince oneself that the work on the right endcap can be found merely by reversing the sign on $\b$.  The net work done against radiation pressure is consequently
\be
\D W = \D W_+ - \D W_- = \frac{4}{3}\left[\frac{4\pi i_o A_o D}{c}\right]\frac{v^2}{c^2}.
\ee
From Eq. (\ref{ro}), however, we recognize that the quantity in brackets is just $\r_o V = E_o$ and so, finally,
\be
\D W = \frac{4}{3} E_o \frac{v^2}{c^2},
\ee
exactly Hasen\"ohrl's result and as predicted in \S\ref{sec2} above by considerations of the energy-momentum tensor.  It is noteworthy that this calculation avoids any integration over an extended hypersurface and this is  because we have effectively ignored the cavity and merely calculated the work on the two radiators.  Consequently, it would be difficult to argue that we have performed some non-covariant procedure and, therefore, modulo any algebraic mistakes, it would appear that this part of Hasen\"ohrl's calculation is consistent with relativity.

\subsection{Relativistic calculation of the energy}
\label{subsec3.3}

We can now use a similar procedure to calculate the energy filling the cavity, which amounts to integrating Eq. (\ref{dEdt}).  Using Eqs. (\ref{i'th'}) and (\ref{dt'}) for $i'$ and $dt'$ respectively, we have
\bea
dE'_{\g A}& =& \frac1{c}\int_{\th'}\int_{t'} {i'} (\cos\th' - \b)\, d \O'\, dt'\, A_o\nonumber\\
 &=& \frac{i_o A_o D'}{c}\int_{\th'} \frac{(1-\b^2)^2}{(1-\b\cos\th')^4}  d \O'.
\eea
With $d\O' = -2\pi d(\cos\th')$ and expressions (\ref{dV'}) and (\ref{i'th'}) from above,
\be
E'_{\g A}= \frac{2\pi D A_o i_o (1-\b^2)^{5/2}}{c}\int_0^1 \frac{d(\cos\th')}{(1-\b\cos\th')^4}.
\ee
(Technically, the limits of integration should be such that $0 \le \th \le \pi/2$ in the cavity frame, which by Eq. (\ref{costh'}) means the limits in the lab frame are $[1, \b]$, but this makes no difference to the required precision of the calculation.) As before, changing the sign of $\b$ gives the contribution from the right endcap. Thus, since the denominator is well behaved at $\cos\th'=0$, the total energy will be
\be
E'_\g = \frac{2\pi D A_o i_o (1-\b^2)^{5/2}}{c}\int_{-1} ^1 \frac{d(\cos\th')}{(1-\b\cos\th')^4}.
\ee
One can work this out to second order in $\b$, but the  integral is easily evaluated analytically with the result
\be
E'_\g = \frac{4\pi D A_o i_o (1-\b^2)^{5/2}}{3c}\left[\frac{3+\b^2}{(1-\b^2)^3}\right].
\ee
Now working to second order in $\b$ gives
\be
E'_\g = E_o (1+ \frac{5}{6}\b^2),
\ee
in agreement with the above result (\ref{E'3}) achieved by transforming the energy-momentum tensor.  Notice that, as in the calculation of the work, we have not performed any integration over an extended spatial hypersurface, and hence it is difficult to see any noncovariant operation in this procedure that violates the principles of relativity.  Furthermore, due to the agreement with the earlier result, there can be little doubt that the energy of the radiation has increased by $(5/6)E_o\b^2$ and that somehow we are missing $(1/2)E_o\b^2$ worth.  We haven't yet discovered how to account for the missing energy, but the calculation just carried out reveals generally why Hasen\"ohrl did not obtain the correct result: \emph{although calculation of the work requires computations to only first order in $\b$, calculation of the energy requires computations to second order in $\b$, where relativistic corrections are important.}  Invoking the work-energy theorem to equate the work to the classical kinetic energy is evidently illegal, but this only deepens the puzzle, since $(1/2)mv^2$ is the correct kinetic energy to second order. Could there be some $p dV$ work we have forgotten?  The only volume change is due to the Lorentz transformation and it is not observed in any given reference frame.  Is there a potential energy in a photon gas?  What is the nature of heat?  Have we discovered a third form of energy that is neither kinetic nor potential?

\section{Cavity stresses?}
\setcounter{equation}{0}\label{sec4}

 Even as desperation mounts, we have not forgotten the similarity between the Hasen\"ohrl cavity and the classical electron.  The proper resolution to that dilemma has resulted in a not entirely civil controversy now lasting over a century.  Left unexplained in the above discussion is why the classical electron, consisting of mutually repulsive charge elements, should be stable.  For this reason in 1906 Poincar\'e introduced\cite{Poin06} the famous ``Poincar\'e stresses," external and unidentified nonelectromagnetic stresses meant to bind the electron together.   If we temporarily adopt such a proposal we can follow the approach initiated by von Laue\cite{Laue11} in 1911 and take the total stress-energy tensor of the system to be
 \be
 T_{\m\n} = (T_{\m\n})_{em} + (T_{\m\n})_{mech}, \label{TT}
 \ee
whose divergence  $T_{\m\n}\,^{,\n}=0$.  Now, for a frame moving at velocity $\bf v$ in the $- x$ direction with respect to the cavity (zero-momentum) frame we have by definition
\be
 T'_{\m\n} = \L_\m \,^\a  \L_\n\,^\b T_{\a\b},  \label{TL}
\ee
with Lorentz transformation matrix components $\L_0\,^0 = \g$, $\L_0\,^x = -\b\g$ and $\L_x\,^x = \g$.  Thus  \be
T'_{00} = \g^2 T_{00} + \g^2 \b^2 T_{xx} + 2\g^2 \b T_{0x} \label{T'00}
\ee
and
\be
T'_{0x} = \g^2 \b T_{00} + \g^2 \b T_{xx} + \g^2 (1+ \b^2) T_{0x}. \label{T0x}
\ee
Provided that the $T_{\m\n}$ are time independent, $\dpr T_{ij}/\dpr x_j = 0$.  Multiplying this expression by $x$ and integrating by parts, with the additional assumption that $T_{ij}$ is spatially bounded, shows that the volume integrals of the second terms in Eqs. (\ref{T'00}) and (\ref{T0x}) vanish identically. The volume integrals of the momentum components, $T_{0k}$, vanish by definition in the zero-momentum frame.  Thus, the remaining first terms give for energy $E'$ and momentum $G'$
\be
E' = \g^2 \int T_{00} dV' = \g \int T_{00} dV_o = \g E_o
\ee
and
\be
G' = \frac{\g^2 \b}{c} \int T_{00} dV' = \frac{\g \b}{c}\int T_{00} dV_o = \frac{\g v}{c^2}E_o,
\ee
where we have used Eq. (3.5).  Consequently, the energy and momentum of an extended body defined in this way transform as components of a 4-vector, consistent with identifying the mass of the closed system to be $E/c^2$. This is von Laue's theorem. (Klein\cite{Klein18} extended Laue's proof to time-dependent, closed systems.)\\

Poincar\'e suggested more than one model for stabilizing stress\cite{Cuvaj68} in a way consistent with these ideas. von Laue and Klein assure us that any model for the stress will suffice so long as the divergence of the total stress-energy tensor vanishes. Here let us assume that the electron is an elastic sphere of  radius $a$ with  constant density $\r$ and pressure $p$, and a uniformly distributed charge on the surface.  The contribution to the energy and momentum from $\r$ is just $\g m_o c^2$ and $\g m_o v$, respectively, and  does not interest us.  On the other hand, the $T_{kl}$ component of the stress-energy tensor of the electron's electromagnetic field is, as usual,
\be
T_{k\ell} = \frac1{4\pi}\left[-(E_k E_\ell + B_k B_\ell) + \frac1{2}({\bf E^2 + B^2)}\d_{k\ell})\right]. \label{Tkl}
\ee
In the electron rest frame we assume that the electric field is given by the Coulomb field ${\bf E} = (e/r^2) { \bf \hat r}$ for $r > a$; ${\bf E} = 0$ for $r < a$ and ${\bf B} = 0$ everywhere.
As shown in \S\ref{subsec3.1} this yields for the total energy of the field $U_o = e^2/2a$.  The requirement that the divergence of the total stress-energy tensor vanish everywhere is satisfied if the Poincar\'e pressure, $(T_{rr})_{mech}$, exerted on the inside surface of the sphere is equal to the electromagnetic pressure, $(T_{rr})_{em}$, on the outside surface of the sphere. From Eq. (\ref{Tkl}) the latter is just
\be
(T_{rr})_{em} = (T_{rr})_{mech}= -\frac1{8\pi} \frac{e^2}{a^4}= - \frac{U_o}{4\pi a^3}.
\ee
From Eq. (\ref{Tuv}) or (\ref{TL}) we see that in the primed frame this results in a  mechanical contribution to the energy density of
\be
(T'_{00})_{mech} = -\g^2\b^2\frac{U_o}{4\pi a^3}.
\ee
Integrating over the volume of the sphere with $V' = V/\g$ gives for the total mechanical energy $E'_{mech} = -\g\b^2 U_o/3$.  We already know from \S\ref{subsec3.1} that transforming $(T_{00})_{em}$  yields $U'_{em} = \g (U_{o})_{em} (1+\b^2/3)$.  Thus the new mechanical term kills the unwanted term in $U'_{em}$, yielding $U' = \g U_o$, as fervently desired.  Similarly, the infamous 4/3 in the expression for $G'$ becomes unity.\\

Because the stress-energy transformation equations for blackbody radiation are identical to those for the electromagnetic field, it is not surprising that commentators, for example the authors of the Wikipedia entry on Hasen\"ohrl mentioned in the Introduction, have jumped to the conclusion that the resolution to the Hasen\"ohrl dilemma and the electron is the same.  Indeed, Hasen\"ohrl at no time in H1, H2 or H3 discusses the physics of the cavity walls, other than to say that they are perfectly reflecting, and so  his error must have been neglect of the cavity stresses.  Such haste is merely a result of ``telegraph physics," and shows that the authors have only repeated what they have heard, rather than having read the original papers.  Cunningham\cite{Cunn1914}, cited by Herrmann\cite{Hermann04} in the Introduction as having corrected Hasen\"ohrl's error, is in fact  concerned with the classical electron as above and never mentions Hasen\"ohrl.   Yet even a cursory examination of H1 reveals three fatal differences between the  electron problem and Hasen\"ohrl's cavity.

The first is that there is no radiation pressure in the classical electron problem; there is in Hasen\"ohrl's.  The electron is not radiating and it will appear to move at a constant velocity in any inertial frame without the imposition of external forces. As Rohrlich points out\cite{Rohr60},  Poincar\'e stresses are introduced both to a) glue the electron together and b) to ensure the Lorentz covariance of the theory.  They are not forces that push the electron.  This is obvious if you are a member of the Fermi school (\S\ref{sec6} below).  You ignore the question of the stability of the electron and  argue that Poincar\'e stresses are merely ``fictitious" forces arising from an improper, noncovariant treatment of the problem.  If  calculations are performed correctly,  Poincar\'e stresses simply do not exist.  Clearly, then, they cannot be the external forces that Hasen\"ohrl requires to keep the cavity moving at a constant velocity against radiation pressure.

The second fatal difference is that internal stresses cannot provide external forces.  Regardless of whether one regards Poincar\'e stresses as genuine or imaginary, the radiation in Hasen\"ohrl's experiment must unquestionably be contained in a real cavity; therefore cavity stresses do arise that must balance the radiation pressure $\r/3$. Relativistically such stresses, as we have seen,  transform  as any other form of energy and should be taken into account.  In Hasen\"ohrl's cavity, however,  only    azimuthal stresses are present, because the external forces $\bf F$ acting against the radiation pressure obviate the need for longitudinal ones.  Longitudinal cavity stresses in Hasen\"ohrl's problem are zero and cannot be enlisted as a solution.

The third fatal difference between the Hasen\"ohrl cavity and the electron is one of symmetry.  The electron is spherically symmetric and has ${\th\th}$ and ${\phi\phi}$  stress components, which must be included when transforming the stress-energy tensor.  However, once again
$ T'_{\m\n} = \L_\m \,^\a  \L_\n\,^\b T_{\a\b}$.  Hasen\"ohrl's cavity is cylindrically symmetric and  moving solely in the $x$-direction; hence as stated below Eq. (\ref{TL})  the only nonzero $\L$'s are $\L_0\, ^0$, $\L_0\, ^x$ and $\L_x\, ^x$.  Any azimuthal stresses, which will have $y$- and $z$-  components alone, simply do not transform into longitudinal forces and  cannot be invoked to resolve the dilemma.

\section{Rocket equation}
\setcounter{equation}{0}\label{sec5}

\subsection{Constant velocity case}
\label{subsec5.1}

With cavity stresses ruled out, one racks one's brains to find the missing piece of physics in order to close the discrepancy between the  calculations of the energy and of the work.  Perhaps one must consider gravity...As darkness closes in, one recollects two crucial facts from the first days of freshman physics: The first is that for two bodies $A$ and $B$ in contact, an external force  $\bf F$ exerted on $A$ is not in general equal to the force $A$ exerts on $B$. The second is that  $\bf F$ is not equal to $m\bf a$, but to $d{\bf p}/dt$.  Modern texts invariably restrict themselves to the  relativistic extension of the former case, $f_\m = m_o (du_\m/d\t)$, for four-force $f_\m$.  Then, if $\bf F$ is the Newtonian three-force,
\be
\bf f \equiv (\frac{\g}{c} { F \cdot v}, \g  F), \label{4f}
\ee
and the four-velocity and the four-force are orthogonal: $f_\m u^\m = 0$.

But these expressions are only valid when the rest mass is constant. If not, one must employ the proper  relativistic definition of force:
\be
f_\m = \frac{dg_\m}{d\t}
\ee
for four-momentum ${\bf g} \equiv (\g m_o c, \g m_o \bf v)$.

In Hasen\"ohrl's experiment, the radiators $A$ and $B$ are filling the cavity with radiation.  They must therefore be losing mass.  In that case, the invariant scalar product of $f_\m$ and $u_\m$ is
\be
f_\m u^\m = u^\m \frac{d(m_o u_\m)}{d\t} = m_o  u^\m \frac{du_\m}{d\t} + u^\mu u_\m\frac{d m_o}{d\t} = \frac1{2} (u^\m u_\m)^{\bf \centerdot} \, m_o  -\dot m_oc^2 = -\dot m_o c^2,  \label{mc2}
\ee
where $(\cdot) \equiv d/d\t$.  Thus $f_\m u^\m = 0$ \emph{only when mass is constant}.

Furthermore, only in the constant-mass case is the time component of Eq. (\ref{4f}) equal to the power $\g \bf F \cdot v$ expended in moving the particle.  If the mass is gaining a quantity of heat $Q$, then the heat gain  also contributes to $dE/dt$ and the time component must be modified to include the rate of heat transfer:
\be
f_0 = \frac{\g}{c} ( {\bf F \cdot v} + Q_{,t}),  \label{f0}
\ee
where $Q_{, t} \equiv dQ/dt$ (\emph{note: dt, not d$\t$}). Then  $f^\m u_\m = -\g^2 Q_{,t}$.  Since this must also be an invariant, we have from Eq. (\ref{mc2})
\be
\g^2 Q'_{,t} = \dot Q_o = \dot m_o c^2, \label{Qt}
\ee
as expected. From these expressions it follows immediately that the total amount of heat transferred in the primed frame is
\be
\D Q' = \g^{-1} \D Q_o.
\ee

This transformation law is at first unsettling, because one would think that the energy in the moving frame should be increased by a factor of $\g$ over the rest energy.  The important point is that $\D Q'$  itself is not the total change in energy, which is given by integrating Eq. (\ref{f0}).  With  $f_0 \equiv dg_0/d\t = (\g/c) dE/dt$,
\be
 \frac{dE}{dt} = ( {\bf F \cdot v} + Q_{,t}).
\ee
Because for constant velocity,  $f_k= \g F_k = \dot m_o u_\k   = (\dot Q_o/c^2) \g v_k$, the three-force is  given by
\be
 F_k = \frac1{c^2} \frac{dQ_o}{d\t} v_k.  \label{Fk}
\ee
Consequently,
\bea
\frac{dE}{dt} &=& ( \dot Q_o \frac{v^2}{c^2} + Q'_{,t})\nonumber\\
                &=& \dot Q_o \frac{v^2}{c^2} + \g^{-2}\dot Q_o \nonumber\\
                &=& \frac{dQ_o}{d\t},
\eea
and the total change in energy in the primed frame is
\be
\D E' = \int \frac{d Q_o}{d\t} dt = \g \D Q_o,
\ee
which is merely the total heat radiated by the endcaps in the rest frame of the cavity, multiplied by $\g$, as desired.  (For a complete, if rococo, discussion see M\o ller\cite{M62}.)\\

\begin{figure}[htb]
\vbox{\hfil\scalebox{.6}
{\includegraphics{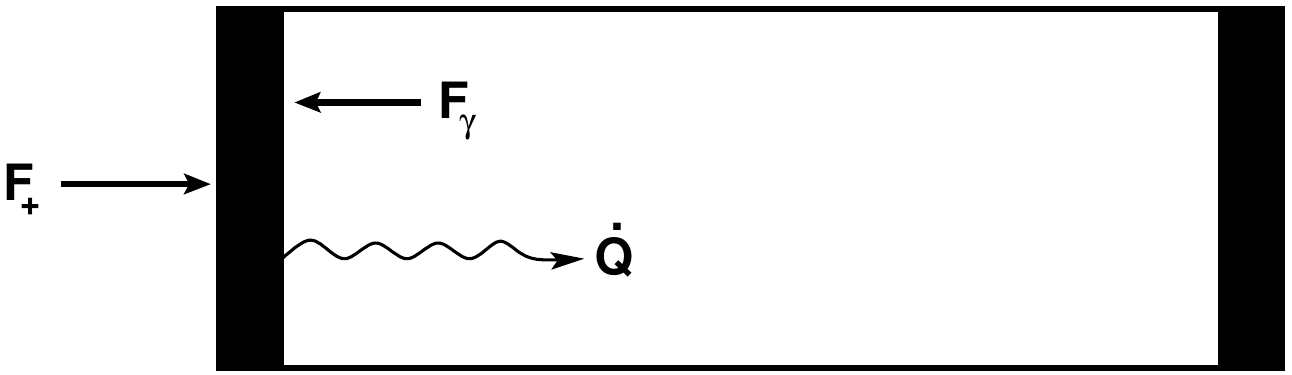}}\hfil}
{\caption{\footnotesize{The external three-force $F_+$ is applied to the endcap to the right.  The reaction force of the radiation is to the left.  However, we must take into account the mass loss of the endcap, which fills the cavity with radiation. \label{cav2}}}}
\end{figure}

With these preliminaries, one can readily make Hasen\"ohrl's thought experiment compatible with relativity.  Referring to figure \ref{cav2} and Eq. (\ref{Fk}), we see that  the total three-force on the left endcap will be
\be
{\bf F = F_+ + F_\g} = \frac{\g}{c^2} \frac{dQ_o}{dt} \bf v.
\ee
Now, taking positive to the right and letting $Q_o$ represent the heat lost from both endcaps, the \emph{external} work on the system will be
\be
\D W_{ext} = \int F_{ext} v dt =  \frac{\g v^2}{c^2}\int \frac{dQ_o}{dt} dt +\int F_\g v dt.  \label{DWQ}                    \ee
From the results of \S\ref{subsec3.2} we know that the second term on the right is  $4/3E_o \b^2$, which is the total work performed \emph{on the radiation}. The first term on the right is merely $\g\b^2\D Q_o$, which must be $-\g\b^2 E_o$, since the endcaps supplied all the radiation in the cavity, $E_o$.  Therefore, contrary to Hasen\"ohrl's result we find to the required order
\be
 \D W_{ext} = \frac{1}{3}E_o\b^2.
\ee
It is easy to see that this value of the work is consistent with the work-energy theorem.  Recall from \S\ref{subsec3.3} that $\D E'_\g \approx (1+(5/6) \b^2)E_o$.  The change in energy of the caps is $\D E'_{caps} = -\g E_o \approx -(1+(1/2)\b^2)E_o$.  Then $\D E'_{caps} + \D E'_\g = + (1/3)\b^2 E_o$, as just found; energy is conserved. However, note that the work-energy theorem does not hold term by term.    This is evidently a consequence of the fact that we have divided the system into two separate energy-momentum tensors, as in Eq. (\ref{TT}) above, whose integrals individually do not behave covariantly. We return to this issue below, after considering acceleration.\\

\subsection{Slowly accelerating case}
\label{subsec5.2}

As we have seen, Hasen\"ohrl carries out his calculation for one light-crossing time.  Once photons from either side strike the other, no further work is performed.  Moreover, we have assumed in all our calculations thus far, he takes the velocity to be constant.

In H2, while not disavowing such an approach, Hasen\"orhl does state  that during the light-crossing time the cavity must have been accelerating.  Large accelerations would result in nonequilibrium, irreversible processes and be impossible to treat, but  quasi-static, infinitesimal changes in velocity might be tractable. Hasen\"ohrl undertakes such a calculation in H2 and one can guess that the total work  should be half that of the constant velocity case.  This is essentially the origin of the factor of two change that appears in the correction paper, H3.   Initially, however, Hasen\"ohrl  gets the same result as in H1, $m = (3/8)E_o/c^2$, until Abraham points out a simpler way to calculate the mass, as the derivative of the electromagnetic momentum with respect to velocity: $m = d[(3/4)E_o v/c^2]/dv = (3/4)E_o/c^2$.  Hasen\"ohrl then uncovers a factor of two error in H2, which brings him into agreement with Abraham. For this reason, some commentators have said that Hasen\"ohrl ultimately did nothing different from Abraham, but Abraham was considering the classical electron, while Hasen\"ohrl was considering electromagnetic radiation.  Only the ``4/3,"   present in both systems, has led to this erroneous conclusion.

Hasen\"ohrl's calculation in H2 is extremely involved.  He does not calculate the work directly, but rather calculates the  infinitesimal change in energy of an apparently already filled cavity due to an incremental change in velocity, then subtracts  the fraction he believes is  emitted by the radiators, leaving the fraction he counts as kinetic energy, which he then equates to the work.  We can more simply obtain a result as follows.  If we regard the cavity radiation as a relativistic photon gas, then the external force density (force per unit volume) on this gas will be given by $\mathfrak  f_\m = T_{\m\n}\, ^{, \n}$, or from Eq. (\ref{Tuv})
\be
\mathfrak  f_\m = \frac{d(\r_o + p_o)}{c^2\, d\t}u_\m + \frac{(\r_o + p_o)}{c^2}\left[\frac{d u_\m}{d\t} + u_\m \frac{\dpr u_\n}{\dpr x_\n}\right] + \frac{\dpr p_o}{\dpr x_\n} \h_{\m\n}.  \label{fm}
\ee
Now, in analogy to the analysis of the previous section, we take the 4-force density to be
\be \vec{\mathfrak f} = [(1/c)({\cal F} \cdot {\bf v} + q,_t), \cal F]
\ee
for Newtonian three-force density $\cal F$ and heat density $q$.  Forming the scalar product $u^\m \mathfrak f_\m = -\g  q_{, t} = -\dot q_o$ allows us to eliminate, among others, the Div$\, \bf  u$ term in Eq. (\ref{fm}) and derive the fundamental equation of motion for the radiation:
\be
\mathfrak f_\m = \frac{(\r_o + p_o)}{c^2}\frac{du_\m}{d\t} + \frac{u_\m}{c^2} \frac{dp_o}{d\t}+ \frac{\dpr p_o}{\dpr x_\n}\h_{\m\n} + \frac{\dot q_o u_\m}{c^2}.  \label{fundeq}
\ee

We now restrict ourselves to the situation $ax/c^2 \sim v^2/c^2  < < 1$, which is equivalent to $(a/c)(D/c)  << 1$, or that the light-crossing time $D/c$ is much less than the  timescale to reach relativistic velocities.  In this case we use the equilibrium values of $\r_o$ and $p_o$ (= $\r_o/3$).  Assuming as usual motion only in the $x$-direction and retaining terms only of order $a \sim v^2/c^2$, one gets for the spatial force density in the lab frame
\be
\mathfrak f' =  \frac{4}{3c^2}\r_o a + \frac{\dpr p_o}{\dpr x}(1 +  \b^2) + \frac{\dot q_o v}{c^2},  \label{fundeq2}
\ee
 The total force on the fluid will then be
\be
F= \int_{V'} \mathfrak f' \, dV' =  \int_{V_o} \left[\frac{4}{3c^2}\r_o a + \frac{\dpr p_o}{\dpr x}(1+  \b^2) + \frac{\dot q_o v}{c^2}\right]\g^{-1} dV_o.  \label{fundeq2}
\ee

Now, in general the time-dilation factor $\g = (-g_{\m\n} dx^\m dx^\n)^{-1/2}$.  From Doppler-shift arguments\cite{Hartle03} or considerations of an accelerating frame\cite{M62,MTW}, one can show that for constant acceleration
\be
\g = \frac1{(1 + \frac{2ax}{c^2} - \frac{v^2}{c^2})^{1/2}}.
\ee
 This may be regarded as a manifestation of the equivalence principle or gravitational redshift.  We have   $\g^{-1} \approx 1 + ax/c^2 - (1/2)\b^2$ and hence to order $a$
\be
F= \int_{V_o} \left[\frac{4}{3c^2}\r_o a + \frac{\dpr p_o}{\dpr x}(1+  \frac{ax}{c^2} + \frac1{2}\b^2) + \frac{d q_o}{dt}\frac{v}{c^2}\right] dV_o.  \label{fundeq3}
\ee
Assuming  $a= dv/dt$, the first term immediately gives Work $= \int F v dt = (2/3)E_ov^2/c^2$, which is Hasen\"ohrl's second result, i.e., setting this term equal to $(1/2)mv^2$ yields $E_o = (3/4) m c^2$.  However, we see that he has effectively neglected the remaining terms in the integral.   If one assumes that $v$ is instantaneously zero in a co-moving frame, then
\be
F= \int_{V_o} \left[\frac{4}{3c^2}\r_o a + \frac{\dpr p_o}{\dpr x}(1+  \frac{ax}{c^2} ) \right] dV_o.  \label{fundeq4}
\ee
To integrate  the second term we make use of  the vector theorem $\int_V (\del p_o)\, dV = \oint_A p_o \,\hat {\bf n}\, dA$ for outwardly directed normal $\hat {\bf n}$.  We integrate the third term by parts with and get
\be
\int_{V_o}  \frac{\dpr p_o}{\dpr x}(1+  ax ) \, dV_o = \oint_A p_o \hat {\bf n}\, dA + \frac{a}{3c^2} (\r_o V_o)|_{boundary} - \frac{a}{3c^2}\int_{V_o} \r_o \frac{\dpr x}{\dpr x} dV_o.  \label{fundeq5}
\ee
Assuming the density and pressure are spatially bounded we can  extend the integration volume to infinity and  drop the boundary terms, which leaves only the last term.  Of course, $\dpr x/\dpr x = 1$ and from Eq. (\ref{fundeq4}) the total force consequently becomes
\be
F = \frac{E_o}{c^2} a.  \label{forcefinal}
\ee
If $F = ma$, we immediately have  $E_o = mc^2$ and the work-energy theorem is not needed.

Note in the above that we did not need to assume that $v$ was instantaneously zero.  The volume integration is independent of $v$ and thus the integral of the term $(1/2) (\dpr p_o/\dpr x)\b^2$ in Eq. (\ref{fundeq3}) vanishes.  Also, in contrast to the constant-velocity case, here we are evidently  assuming that the cavity is already filled with radiation.  Then from \S\ref{sec3} the total heat injected into the cavity is $\D Q_o \sim E_o \b^2$. The volume integral of the heat term  is hence $(d( \D Q_o)/dt)  \b \sim (\dpr (\D Q_o)/\dpr x )\b^2 \sim E_o \b^4$ and can be ignored.  Finally, since we eliminated the Div {\bf u} term in favor of the heat term, no assumption has been made about the incompressibility of the fluid.  The result appears completely general at this level of approximation.

One might legitimately ask, however, given that our individual calculations of the energy and the work gave the wrong answers, why one should believe a direct calculation of the force.  From the von Laue-Klein theorem, it is not difficult to demonstrate that an external force $\bf F$ applied to a cavity filled with radiation results (in the limit $v << c$) in an acceleration ${\bf a = F}/(E_o/c^2)$, where $E_o$ is the rest-frame energy of the  radiation plus cavity, and $\bf a$ is the acceleration in the zero momentum frame.  For cavity rest mass $m_c$ this implies that the total work on the cavity plus radiation $\D W_{tot} = (1/2) (m_c + E_{o\g}/c^2) v^2$ consistent with the above result.  Furthermore, we know that the final energy of the radiation is $E_{o\g} (1 + (5/6) v^2/c^2)$ and
the final energy of the cavity is $E_c = ( m_c c^2 + (1/2) m v^2 - (1/3) E_{o\g} v^2/c^2)$
where the last term is due to the cavity stresses.  Thus the final total energy is
$E_f = m_c c^2 + (1/2)m_c v^2 + E_{o\g} (1 + (1/2)v^2/c^2)$, which is what one would naively claim, but as we see, one must be extremely careful when considering the energy of individual parts of the total system.\\

Although our treatment shows how to close the factor of two discrepancy between Hasen\"ohrl's  two calculations,  generally the reason he achieved an incorrect result in the constant-velocity situation is that he wants to rigorously equate the work performed to kinetic energy, as the work-energy theorem demands.  Unfortunately, he does not know how to properly compute the energy.  In particular, Hasen\"ohrl does not conceive of the fact that if the radiators are losing energy, they must be losing mass, which contains an element of irony because it is precisely a mass-energy relation that he is trying to establish. In the slowly accelerating case, as we have just seen, heat transfer is negligible, and one can bypass use of the work-energy theorem (which is just the time integral of $F=ma$) but one needs a correct calculation of the force, which requires relativistic corrections.  In any case, that calculating the mass from acceleration  results in a different answer than calculating it from constant velocity is not surprising and, as we have now shown, both cases can be made consistent with relativity.

\section{Fermi and Hasen\"ohrl}
\setcounter{equation}{0}\label{sec6}

Two of Enrico Fermi's earliest papers, from 1923, are devoted to matters that touch directly on Hasen\"ohrl's thought experiment.  In the first\cite{Fermi22}, Fermi discusses the 4/3 problem for the classical electron and states that the paradox arises because the electron is assumed to be a rigid body, in contradiction to the principles of special relativity.  He then applies the concept of ``Born rigidity" to the electron, which requires that given points in an object always maintain the same separation in a sequence of inertial frames co-moving with the electron. Equivalently, Born rigidity demands that the worldline of each point in the electron should be orthogonal (in the Lorentzian sense) to constant-time hypersurfaces in the co-moving frames (see, eg., Pauli\cite{Pauli}).  However, such constant-time hypersurfaces are of course not parallel to those in the lab:  As acknowledged in \S\ref{subsec3.1}, a constant-time integration over the electron's volume in its rest frame assumes that two points on the electron's diameter cross the $t=0$ spatial hypersurface simultaneously, but this will \emph{not} be the case in a Lorentz-boosted frame.  We will see momentarily how Fermi shows that the postulate of Born rigidity allows the elimination of  the spurious factor of 4/3 in the electron's momentum.  In a second 1923 paper, already mentioned in the Introduction, Fermi and coauthor Pontremoli\cite{Fermi23b} employ the same technique to ``correct" the radiation problem.

Fermi's approach to the electron requires calculation of the self-force, the force that the various parts of an electron exert on each other.  We paraphrase his procedure as follows (for more details see \cite{Jackson75} and \cite{Bini11}).  Assume the Lagrangian $L = (de/\g c)u_\m A^\m$, which represents a charge element of the electron in an electromagnetic field. $A^\m \equiv (\Phi, \bf A)$ is the four-potential and we let $de = \r dV$ for density of the electron $\r$.  The action then becomes
\be
S =\frac1{c} \int_V\int_{t}  \frac{\r}{\g} u_\m A^\m dV\, dt = \frac1{c} \int_V\int_{x_\m} \r\, dx_\m A^\m dV.
\ee
Regarding the $A^\m$ as a function of the $x_\m$, the usual variational procedure results in
\be
\d S = \frac1{c} \int_V\int_{x_\m} \r F^{\m\n} dV \, dx_\m  \d x_\n = 0,
\ee
where $F^{\m\n} \equiv \dpr A^\m/\dpr x_\n - \dpr A^\n/\dpr x_\m $ is the electromagnetic field tensor.  Since the $\d x_\n$ are arbitrary and the spatial coordinates $x_k$ are functions of time only, we can write
\be
 \d S^\n   =\frac1{c} \int_V\int_{t_1}^{t_2}\r\, F^{\m\n}  \, \frac{dx_\m }{dt} dV \,dt = 0. \label{dS}
\ee
For simplicity, Fermi chooses the electron to be instantaneously at rest, in which case only the $F_{0k} = E_k$ terms survive.  Then the variation boils down to the vector equation
\be
\int_V \r {\bf E}\, dV = 0.
\ee
To calculate the self-force, one must decompose the electric field and vector potential into external fields and self-fields: ${\bf E} ={\bf E}_{ext} + {\bf E}_{s}$, in which case
\be
 \int_V \r  {\bf E_{ext}} \, dV + \int_V \r {\bf E_{s}}\, dV = 0.
\ee
Provided that the external field is roughly constant over the size of the electron, ${\bf E_{ext}}$ can be pulled out of the integral and the  first term can be identified with ${\bf F_{ext}}$.  We already know from Eq. (\ref{G'}) that the momentum ${\bf G} \approx (4/3)U_o {\bf v}/c^2$, which implies that ${\bf F}_{ext} = d{\bf G}/dt  \approx (4/3)U_o {\bf a}/c^2$ and so we have immediately
\be
\int_V \r {\bf E_{s}}\, dV = -\frac{4}{3c^2}U_o {\bf a}, \label{Es}
\ee
a result that can  be verified by a direct evaluation of the retarded self-fields in the Abraham-Lorentz model of the electron (see \cite{Jackson75}) for gory details).  Also, if one  sets ${\bf F}_{ext} = m\bf a$, one concludes with Abraham that
\be
m =  \frac{4}{3c^2}U_o,
\ee
as already mentioned in \S\ref{subsec5.2}, and which explains why Hasen\"ohrl was satisfied with 4/3.

\begin{figure}[htb]
\vbox{\hfil\scalebox{.6}
{\includegraphics{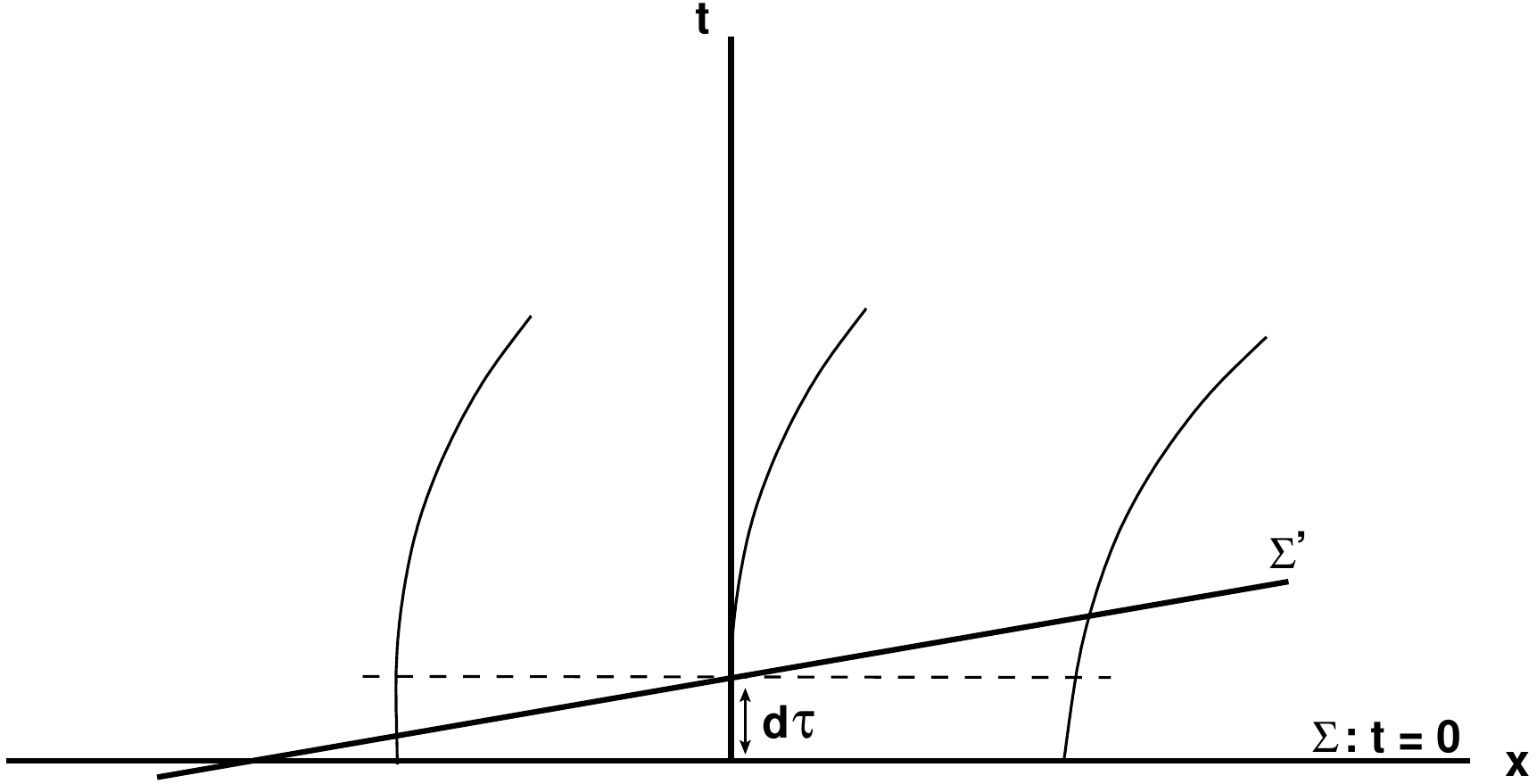}}\hfil}
{\caption{\footnotesize{The worldlines of different points on an accelerating electron will not be straight lines and will intersect  spatial hypersufaces $\Si'$ at different times.  The postulate of Born rigidity assumes that the worldlines intersect the hypersurfaces orthogonally in the Lorentzian sense. (From Bini et. al\cite{Bini11}.) \label{ferm1}}}}
\end{figure}

Fermi, however, rejects the above procedure.  Figure \ref{ferm1} shows the worldlines of several points on an accelerating electron.  We assume that the $x$-axis, $\Si$, represents a $t = \t = 0$ spatial hypersurface.  For an electron instantaneously at rest in a co-moving frame, the time interval along the central worldline will be approximately the proper-time interval $d\t$ of an observer in that frame.  Note that the $\Si'$ hypersurface has an equation of the form $t\equiv   dt = mx + d\t$, for slope $m$.  By assumption of Born rigidity the worldlines of the electron points are orthogonal to this line (in the Lorentzian sense $\h_{\m\n}dx^\m dx^\n = 0$).  Moreover, for constant $a$ the motion is hyperbolic: $x^2 - c^2t^2 = c^4/a^2$, which implies that the slope of the central worldline is $dt_c/dx_c =1/v_c$ and hence for the hypersurface $\Si'$, $m = v_c /c^2\approx (a/c^2) dt$,  where $a$ is the constant proper acceleration of the central worldline.  Therefore to first order
\be
dt = (1 + \frac{ax}{c^2})d\t.
\ee
(The result can also be derived rigorously as a consequence of ``Fermi normal coordinates," which Fermi developed in an earlier paper\cite{Fermi23a}.  See also MTW, chapter six \cite{MTW}.)

Eq. (\ref{dS}) now becomes
\be
 \d S^\n   =\frac1{c} \int_V\int_{\t_1}^{\t_2}\r\, F^{\m\n}  \, \frac{dx_\m }{dt} dV \,d\t (1 + \frac{ax}{c^2}) = 0,
\ee
or for small $d\t$ and  the same assumptions as before
\be
\int_V \r (1 +  \frac{ax}{c^2}) {\bf E}\, dV = 0.  \label{Et}
\ee
Once again we decompose $\bf E$ into the external and self-fields, which gives
\be
\int \r {\bf E}_s \, dV + \frac{a}{c^2} \int \r\, x {\bf E}_s \,dV + {\bf E}_{ext}\int \r \,dV + {\bf E}_{ext} \frac{a}{c^2}\int \r \, x \, dV =0.
\ee
 Since $\bf E_{ext}  \sim \bf a$, the last term is $\sim a^2x/c^2 \sim v^2/c^2 << 1$ and can be neglected compared to the third term.  With Eq. (\ref{Es}) we then have
\be
{\bf F}_{ext} = \frac{4}{3c^2}U_o {\bf a} - \frac{a}{c^2} \int \r\, x {\bf E}_s \,dV.
\ee
For a spherically symmetric charge distribution the only exceptional direction in the integral is the $x$-direction, and so the $y$- and $z$-integrals vanish.  If ${\bf E}_s = \int (\r' dV'/r^3) \bf r$, where $ \bf r = x - x'$, then $E_x = \int \r' [(x-x')/r^3] dV'$.  However, $x$ and $x'$ are merely dummy variables.  By interchanging  them and adding the terms in the above integral r the $x$-component of the force becomes
\be
F_{ext} = \frac{4}{3c^2}U_o {a} - \frac{a}{c^2} \int\int \r\r'\, \frac{(x-x')^2}{2r^3}\,dV \,dV'
\ee
The integral can either be evaluated directly in spherical coordinates or by recognizing that because all orientations are equally probable in a spherically symmetric system, one can replace $(x-x')^2$ by its average, which is $1/3 r^2$.  Then
\be
 F_{ext} = \frac{4}{3c^2}U_o a - \frac1{3}\frac{a}{c^2} \int\int \frac{\r\r'}{2r} \,dV \,dV'.
\ee
But the integral is precisely $U_o$, and so
\be
F_{ext} = \frac{4a}{3c^2}U_o  - \frac1{3}\frac{a}{c^2}U_o,
\ee
which eliminates the unwanted 4/3. \\

In a second 1923 paper\cite{Fermi23b}, Fermi and Pontremoli (FP) apply  an analogous prescription  to solve the cavity-radiation problem.  They consider the forces applied to a volume of radiation without any mention of heat transfer and, as we did above, restrict their attention to the slowly accelerated case $a^2x/c^2 \sim v^2/c^2 << 1$.  Unfortunately, the authors provide little or no explanation for a number of their statements, and so what follows contains a certain amount of interpretation.

FP first state (in other notation) that the total force on the radiation is
\be
F_{ext} = \int (df_s + df_a), \label{FF}
\ee
where $df_s$ is the force element due to the static radiation pressure and $df_a$ is the force element due to ``perturbations," i.e., acceleration.  This term is taken to be the time derivative of the electromagnetic momentum, or as above,
\be
F_a = \frac{4}{3c^2}E_o a. \label{FFa}
\ee
FP then claim that Eq. (\ref{FF}) is incorrect as it stands and due to the ``notion of rigidity" discussed in the electron paper needs to be modified to read
\be
F_{ext} = \int (df_s + df_a)(1+\frac{ax}{c^2}). \label{FF2}
\ee

By these statements the authors apparently mean that if ${\bf p}_{\g s}$ is the static radiation pressure then from Eq. (\ref{FFa}) one can then write to first order in $a$
\be
F_{ext} = \oint_A {\bf p}_{\g s} \cdot  \hat {\bf n}\, dA + \frac{a}{c^2} \oint_A x {\bf p}_{\g s} \cdot  \hat {\bf n}\, dA  + \frac{4}{3c^2}E_o a.
\ee
This equation resembles Eq. (\ref{fundeq4}) above.  FP assert that the first term vanishes and obtain the same result as we derived earlier: $F= (E_o/c^2) a$. \\

Did Fermi solve the cavity-radiation problem?  One thing seems clear: even in 1923 scientists cited previous work without having read it.  Obviously, the FP solution requires that the cavity be accelerating, which is not the case in  Hasen\"ohl's first thought experiment, and to that situation the FP approach is irrelevant.  Moreover, although FP obtained a correct result,  they did not begin with a full or proper version of the fundamental fluid equation and one must regard their argument as more heuristic than rigorous.  Ultimately, it is unconvincing to base a proof of $E=mc^2$ on Born rigidity, which is not a basic principle; certainly mass-energy equivalence must hold for non-Born-rigid bodies.

In this regard, however, the assumption of Born rigidity results in \emph{precisely} the same correction term term in the equation of fluid motion as does the change in $\g$ due to acceleration.  This is evidently not coincidental.  The requirement of Born rigidity is that the worldlines of two nearby particles remain parallel in a co-moving frame, in other words, experience no tidal forces, which is exactly what the principle of equivalence would demand in a uniformly accelerated frame.  To that extent, it is not surprising that the two methods give the same answer, although at higher orders of approximation they would not.

We observe that Fermi's approach appears to be diametrically opposed to that of von Laue, which we followed earlier in solving the cavity problem. Fermi might argue that relativity prescribes covariant methods for carrying out integrations and that our approach (at least for the accelerating case where we introduced the volume element $\g^{-1}dV_o$) has been meaningless, since an integration over constant-time hypersurface in one frame does not correspond to the same physical events in another frame.  Furthermore, he would point out that in the electron problem the origin of the Poincar\'e stresses is totally unexplained.

One can reply that Fermi provides no explanation for the stability of the electron or, equivalently, how the gas in contained in the cavity.   By providing external stresses,
von Laue's approach answers this question.  Moreoever, as mentioned in \S\ref{subsec5.2}, von Laue's  theorem implies that ${\bf a \approx F}/(E_o/c^2)$, where $E_o$ is the rest frame energy of the  radiation plus cavity.  In the case where the internal energy of the cavity is insignificant, we recover  Eq. (\ref{forcefinal}) and Fermi's result. (This is not to say that one needs to include cavity stresses in the calculations; nowhere did we do so in the derivation leading to $E_o = mc^2$ for the accelerating case.) Lastly, an observer in a given Lorentz frame generally measures lengths and densities at constant time, we know how to transform from one frame to another and von Laue's theorem ensures that we will get the correct answer, even if  stress tensors of each subsystem ($\int (T_{\m\n})_{em}$ and $(\int T_{\m\n})_{mech}$) individually transform in a noncovariant manner.

Such matters remain controversial and, although Fermi's papers were largely forgotten, their philosophy has been taken up over the decades by a number of opponents of the Laue school. The most visible these has probably been Rohrlich\cite{Rohr60,Rohr82} and its most vociferous (and amusing) Gamba\cite{Gamba66}.  More recently Bini et al. have jumped into the Fermi camp\cite{Bini11}.   Prominently taking up the banner of  the Laue school has been Boyer\cite{Boyer82}.  A short review by Campos and Jim\'enez\cite{Campos86} attempts to moderate, although in a more recent paper\cite{Campos08} the authors seem to have defected to the Laue camp.  The second edition of Jackson's \emph{Classical Electrodynamics}\cite{Jackson75} discusses both approaches.  The reader who wishes to enter the fray is referred to these works, but we have demonstrated how to get an answer by both methods and will say no more on the subject. \\

\section{Hasen\"ohrl and Einstein}
\setcounter{equation}{0}\label{sec7}
On learning about Hasen\"ohrl's calculation, the first question from any physicist is inevitably, ``Did Einstein know about it?"  There is, to our knowledge, no smoking gun that provides a permanent answer to this question.  Einstein was not in the habit of citing others in his early papers, a lapse which has given rise to countless myths and speculations about what he knew and what he didn't.  On the one hand it seems unlikely, if not incredible, that he would not have known of an award-winning paper that appeared in the \emph{Annalen der Physik}, the leading physics journal of the time, to which he had himself already contributed five papers before 1905.  On the other hand, he always insisted on his priority in the matter.  In  a 1907 paper Johannes Stark had credited Planck with $E=mc^2$.  Einstein replied  testily, ``I find it rather strange that you do not recognize my priority in the relationship of inertial mass and energy," to which Stark responded contritely that upon rereading Planck he realized that the latter's work was rooted in Einstein's own, at which Einstein himself apologized for his pettiness\cite{Ein_coll}.

Max von Laue always gave Einstein credit, reacting to L\'enard's article\cite{Lenard21} by admitting that Hasen\"ohrl might have made the first attempt to construct a dynamical theory of cavity radiation but, ``But that \emph{every} energy flow carries momentum and that conversely \emph{every} momentum implies a flow of energy is an insight which only the theory of realtivity could reach in a consistent way; for only this theory shattered the foundations of Newtonian dynamics\cite{Jammer00}."  He also rejected L\'enard's proposal to call the inertia of energy ``Hasen\"ohrlsche Masse" as being misleading because the terms ``inertia of energy" and ``mass" are synonymous.  Certainly, if not in 1904 then at some point, Einstein must have learned of Hasen\"ohrl's work.  Jakob Lamb asks Einstein directly in a 1908 letter\cite{Ein_coll} whether Einstein has read Hasen\"ohrl's  recent paper ``On the thermodynamics of moving bodies\cite{H4}." Einstein's replies provide no answer but the famous photograph of the first Solvay conference in 1911 shows  Hasen\"ohrl and Einstein, separated by  five other illustrious physicists, standing around the same conference table.  One can only imagine the conversations.

Apart from historical questions, it is reasonable to ask what constitutes an acceptable proof of a statement that we now regard as a law of nature, and in what respects was Einstein's demonstration more general than Hasen\"ohrl's.  Unquestionably, Einstein's thought experiment in his famous 1905 paper ``Does the inertia of a body depend on its energy content?\cite{Ein05b}" was the simpler and in choosing a problem that is readily solved he displayed the sagacity of a great scientist.   Einstein's major simplification, aside from the introduction of relativity itself, was that he effectively considered a point mass, one that  emits two bursts of electromagnetic radiation in opposite directions.    In a frame, say the lab, where the mass is initially at rest, it  clearly  remains at rest.  Einstein assumes that the radiation carries away an energy $E_\g$ such that
\be
E_i = E_f + \frac1{2} E_\g + \frac1{2} E_\g, \label{Eo}
\ee
where $E_i$ is the initial energy of the particle and $E_f$ is the final energy.  An observer in  a rocket frame moving with velocity $v$ with respect to the lab will see the radiation Doppler shifted, in the manner of our toy model in \S\ref{sec2}. Here however, Einstein uses the relativistic Doppler shift (Eq. (\ref{dop})), which he had  derived a few months earlier\cite{Ein05a}, in order to get
\bea
E'_i &=& E'_f + \frac1{2} E_\g \left(\frac{1- \b\cos\th}{\sqrt{ 1-\b^2}}\right)+ \frac1{2} E_\g \left(\frac{1+ \b\cos\th}{\sqrt{1-\b^2}}\right)\nonumber\\
&=& E'_f + \frac{E_\g }{\sqrt{1-\b^2}}.  \label{E'o}
\eea
 Subtracting equation (\ref{Eo}) from equation (\ref{E'o}), he argues that $E'-E$ must be a kinetic energy, i.e., $E_i'-E_i = (1/2)m_i v^2$ and that $E_f'-E_f = (1/2)m_f v^2$.  Thus one has immediately
\be
\frac{1}{2}\D mv^2 = E_\g \left(\frac1{\sqrt{1-\b^2}} - 1\right),
\ee
where $\D m \equiv m_i - m_f$.  But  to evaluate this expression he now takes the lowest-order expansion of the square root $(1-\b^2)^{-1/2} \approx 1 + (1/2)\b^2$, which leads at once to
\be
E = \D m c^2.
\ee
At first sight, this is a fairly convincing demonstration, far more straightforward than Hasen\"ohrl's and one that has the undoubted advantage of yielding the correct answer.  However, although Einstein has begun with the relativistic Doppler effect, in the last step he has equated the lowest-order expansion to the Newtonian expression for the kinetic energy. Consequently, one must concede that his derivation is, at best, a low-velocity approximation.

Einstein's assumption of a point particle is also open to criticism.  In considering an extended system Hasen\"ohl was being far more audacious, or reckless, than Einstein, because as we have seen throughout the present paper, treatment of extended objects, which must include stresses, enormously complicates the picture. The fact that Einstein attempted to deal with such concerns in six further papers on mass-energy equivalence may be taken as \emph{prima facie}  evidence that he was unsatisfied with his 1905 demonstration.  According to Jammer\cite{Jammer00},  ``in spite of many strenuous efforts [Einstein] never succeeded in establishing a general proof of the relation, that is a proof without premises that are valid only in special cases."  Ohanian\cite{Ohan09} goes further and argues that all  the subsequent papers contain errors.   The correct demonstration of $E=mc^2$ only came about with the tensor methods introduced by Laue\cite{Laue11} in 1911 and their generalization to time-dependent systems by Klein\cite{Klein18} in 1918. We have not investigated Einstein's later papers, but although in the constant-velocity case we managed to ``finesse"  the cavity problem  by considering only forces on the endcaps in a way consistent with ``point-mass" relativity, it seems highly unlikely that in 1905 Einstein himself would have gotten the correct answer had he  tackled Hasen\"orhl's thought experiment.  Certainly with the methods then on hand he could not have solved the accelerating case.

One might also argue that Hasen\"ohrl did not obtain any meaningful result whatsoever because his calculation was purely classical, while mass-energy equivalence is a relativistic concept.  In this regard it may be helpful to examine one of the many alternative demonstrations that physicists have proposed since Einstein for pedagogical reasons and in order to better comprehend the underlying assumptions (see \cite{Jammer51,Jammer00}).  The proof  is in fact on the current Wikipedia page on mass-energy equivalence, under the heading,  ``First correct demonstration (1905)" by Einstein\cite{Wikime}.  Ironically, it is neither Einstein's nor obviously correct.   The demonstration, based on one by Rohrlich\cite{Rohr90}, involves the same mass as above, which  at $t=0$ emits  two bursts of radiation in opposite directions.  In the lab, the mass remains at rest with  initial  and  final momentum equal to zero.  In the rocket frame the block has initial velocity $\bf v$ and, since it remains at rest in the lab, it must remain moving at $\bf v$ in the rocket frame.  But then momentum conservation  requires that the mass of the block  has changed:
\be
m_f {\bf v} = m_o {\bf v} - {\bf G'_{\g +} -  G'_{\g -}}
\ee
or, in one-dimension,
\be
m_f { v}  = m_o v -  G'_{\g +} + G'_{\g -},
\ee
where $G'_\g$ represents the momentum lost to the radiation.  Although Rohrlich works in a photon picture, with $E_\g=h\n$, from classical electrodynamics we can accept that $G_\g = E_\g/c$.  Then, using the classical Doppler shift to calculate the energy in the rocket frame,
\be
m_f v = m_o v - \frac{E_\g}{2c}(1 + \b) + \frac{E_\g}{2c}(1 - \b),  \label{momc}
\ee
and it follows at once that $E = \D mc^2$.

The only thing this derivation has assumed is  momentum conservation,  the Maxwellian relationship between light energy and momentum and the constancy of $c$. Is relativity even involved?  One can argue that classical electromagnetism is a relativistic theory and so, yes, the derivation requires relativity.  Einstein himself remarked in his $E=mc^2$  paper\cite{Ein05b}, ``The principle of the constancy of the velocity of light is of course contained in Maxwell's equations," which might strengthen such a contention (and, \emph{inter alia}, clarifies why he made this postulate the basis of relativity).  But have we actually assumed the constancy of $c$?  Maxwell regarded his theory as probably valid only with respect to the ether and, at first order, it is not necessary to assume that the speed of light is constant: the demonstration works perfectly well in an ether theory, with $c$ equal to the ``absolute velocity of light" and $v$ equal to velocity of the source relative to the ether.

Our point is that this proof is a low-velocity, point-mass Newtonian approximation, which may or may not hold for relativistic objects.\footnote{Rohrlich's demonstration is itself very similar to an earlier one given by Steck and Rioux\cite{SR83}, which uses the relativistic Doppler shift and momentum conservation to get $E=\D mc^2$.}  If one accepts  validity in a demonstration enlisting only nineteenth-century physics, then it seems inescapable to accord Hasen\"ohrl recognition for establishing the equivalence of mass and radiation.  Although he obtained the wrong coefficient, he clearly stated  in H1 that an ``apparent mass of $(8/3) E/c^2$" is added to the energy of the radiation due to the motion of the cavity.  This was a highly nontrivial conclusion and we do not believe that his result was coincidentally, in any respect, nonzero.  Hasen\"ohrl did, on the other hand,  equivocate in H3, writing that the concept of apparent mass is probably only applicable to quasi-static motion.

Finally, it is often claimed that Einstein's proof superceded all previous attempts, not only in correctness, but in  generality.  It is true that Hasen\"ohrl's demonstration most obviously requires that his radiators  be at a nonzero temperature $T$; otherwise no radiation is emitted and the thought experiment fails.  But Einstein's  experiment also requires radiation to be emitted from the mass and he provides no explanation as to how this occurs.  If the radiation is heat, then it is exactly blackbody radiation; if it is radiation due to radioactive decay, as he implies at the end of his paper, then one should explain under what circumstances decay takes place. If no mechanism to transform mass to energy is available, then  the statement $E=\D mc^2$ holds no content. In any case, Einstein is clearly speaking about electromagnetic radiation, and so it  is difficult to see in what sense his thought experiment is a ``universal" statement about mass and energy.  Gradually it became so, but this was the result of the labors of many physicists.

Let us end by saying that Fritz Hasen\"ohrl attempted a legitimate thought experiment and tackled it with the tools available at the time.  He was working during a transition period and did not create the new theory necessary to allow him to solve the problem correctly and completely.  Nevertheless, his basic conclusion remained valid and for that he should be given credit.  More generally, his {\it gedankenexperiment}  raises similar profound issues as the classical electron, issues that remain controversial to the present.  For that reason alone Hasen\"ohrl's conundrum is worthy of study and, indeed, tackling it greatly extends the resources of any student who believes himself to be a master of special relativity.

\section{Appendix: a closer look at Hasen\"ohrl's calculation.}
\begin{figure}[htb]
\vbox{\hfil\scalebox{.6}
{\includegraphics{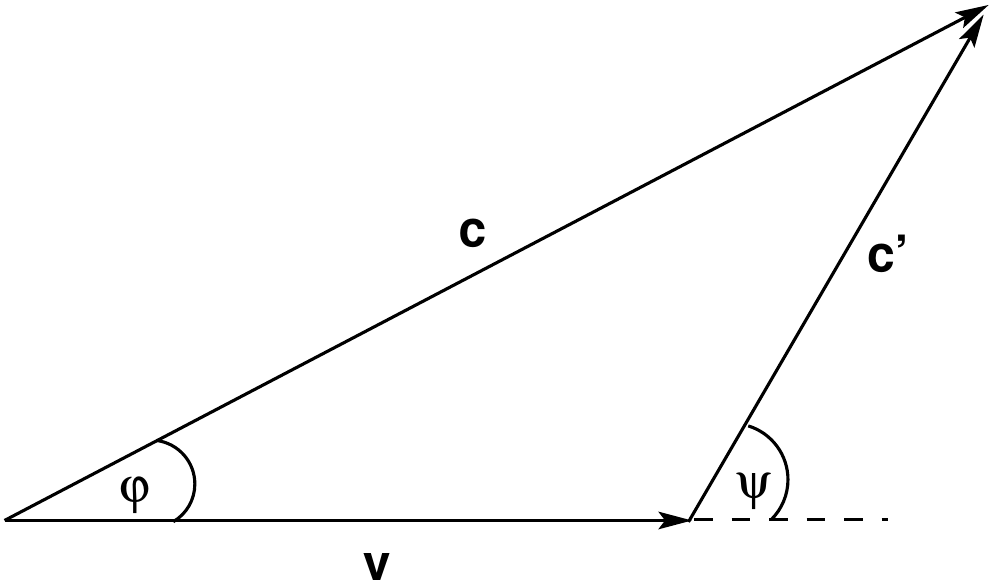}}\hfil}
{\caption{\footnotesize{In Hasen\"ohrl's theory, $c$ is the speed of light with respect to the ether, and $c'$ is the speed with respect to a body moving at velocity $v$.  \ \label{cvecs}}}}
\end{figure}

\setcounter{equation}{0}
\renewcommand{\theequation}{A.\arabic{equation}}
In this paper we have enlisted many modern techniques for solving Hasen\"ohl's problem and it would be a mistake to suppose that he tackled his own thought experiment in the same manner.  We here then briefly outline his major assumptions.  The first is that there is an ``absolute  speed of light" $\bf c $, i.e., the speed of light with respect to the ether,\footnote{ The ``absolute speed of light" is not a constant; in one dimension it is just the sum or difference of $v$ and $c'$.} and a ``relative speed of light" $\bf c'$, the speed of light relative to a body moving at a velocity $\bf v$ with respect to the ether  (figure \ref{cvecs}).
By definition $\bf c = v + c'$ and, if $\ph$ is the angle between $\bf c$ and $\bf v$,  the  ``upstream"   and ``downstream"  velocities of light are respectively
\be
c_\pm = \sqrt {{\bf c' \cdot c'}} = c\sqrt{1 + \b^2 \pm 2\b \cos\ph}. ,
\ee
From the diagram one can easily show that
\be
\cos\ps = \frac{c}{c'}(\cos\ph - \b), \label{cosps}
\ee
and with a bit of trigonometry,
\bea
c_+ &=& c(+\b\cos\ps + \sqrt{1-\b^2\sin^2\ps})\nonumber\\
c_- &=& c(-\b\cos\ps + \sqrt{1-\b^2\sin^2\ps}). \label{cpm}
\eea
From Eqs. (\ref{cosps}) and (\ref{cpm}) it follows directly for $c_-$ that
\be
\cos\ph = \b \sin^2\ps + \cos\ps \sqrt{1 - \b^2\sin^2\ps}, \label{cosph}
\ee
which to first order in $\b$  gives $\cos\ph = \cos\ps + \b\sin^2\ps$, the classical aberration formula. (For quantities relating to $c_+$, one flips the sign on $\b$.)  Since $\ph < \ps $ (source moving in same direction as radiation), we can identify $\ps$ with angles in the cavity frame and $\ph$ with angles in the lab frame, which is fixed to the ether.  Hasen\"ohrl performs his calculations in the cavity frame.

Now, from Abraham he takes the light pressure to be the ``absolute pressure," i.e., measured with respect to the ether:
\be
p_{-} = \frac{i}{c}  \cos\ph, \label{ph}
\ee
where $i$ is the intensity.  Here $p_-$ is what we would call $dp_-$ or specific pressure; in other words, Hasen\"ohrl then calculates  the total pressure as $\int p_- cos\ph d\O$.  Since $\ph$ is in the ether frame and he wants to calculate in the cavity frame, he must transform the angles.  However, he next asserts that the total relative intensity of radiation given off from the endcaps  is
 \be
i = i_o + vp_{-}, \label{ih}
\ee
 where $i_o$ is the same specific intensity that we have used in this paper.  According to this
formula, the intensity of radiation emitted by an endcap in the ether frame, $i$, is the sum of the intensity in the cavity frame, $i_o$, plus the power per unit area expended by the external force to balance the radiation pressure.  From Eqs. (\ref{cosph}) and (\ref{ph}) we have immediately
 \be
 p_{-} = \frac{i}{c}\left[\b \sin^2\ps + \cos\ps \sqrt{1 - \b^2\sin^2\ps}\right], \label{p-}
 \ee
 and inserting this into (\ref{ih}) and solving for $i$ gives
 \be
 i = i_o\frac{c}{c_-\sqrt{1-\b^2\sin^2\ps}}.
 \ee
Using this in Eq. (\ref{p-}) to get $p_{-}$ in terms of $i_o$ yields directly
\be
p_{-} = i_o\frac{\b\sin^2\ps +\cos\ps\sqrt{1-\b^2\sin^2\ps}}{c_-\sqrt{1-\b^2\sin^2\ps}},
\ee
which, after some algebra, can be further rewritten in terms of the absolute $c$ to get
\be
    p_{-} = i_o\frac{\cos\ps + \b\sqrt{1-\b^2\sin^2\ps}}{c(1-\b^2)\sqrt{1-\b^2\sin^2\ps}}.
\ee
Now Hasen\"ohrl has $ p_{-}$ in solely in terms of cavity-frame variables and the absolute speed of light and can use it and the analogous expression for $ p_{+}$ in a calculation of the work, much as we did in \S\ref{subsec3.2}.  As one sees, the classical ether theory results in enormously complicated expressions for the intensity and pressure.  They are, nevertheless, correct to first order in $\b$, which as explained in the text can give the correct result for the work, but not for the energy.  \\

{\bf Acknowledgements}
We are grateful to Thomas Gregor for his translation of some of Hasen\"ohrl's June paper.  We also thank Bob Jantzen for bringing Fermi's papers to our attention, sharing reference\cite{Bini11} before publication, and thank him and Jim Peebles as well for helpful conversations.

{\small

\end{document}